\documentclass[%
 reprint,
 amsmath,amssymb,
 aps,
]{revtex4-2}
\usepackage{xcolor}%
\usepackage{graphicx}
\usepackage{dcolumn}
\usepackage{bm}
\usepackage{float}
\usepackage{verbatim}
\usepackage{amsmath}

\def\Q{\mathbf{Q}}
\def\n{\mathbf{n}}
\def\x{\mathbf{x}}
\def\e{\mathbf{e}}

\def\tr{\textrm{tr}}
\begin{document}

\preprint{APS/123-QED}

\title{Landau-de Gennes  Modelling of Confinement Effects and Cybotactic Clusters in Bent-Core Nematic Liquid Crystals}

\author{Yucen Han}
 \affiliation{Center for Applied Mathematics, Renmin University of China, Beijing 100872, China}
 
\author{Prabakaran Rajamanickam}
\affiliation{Department of Mathematics and Statistics, University of Strathclyde, Glasgow G1 1XH, United Kingdom
\mbox{}}
\author{Bedour Alturki}
\affiliation{Department of Mathematical Sciences, College of Science, Princess Nourah bint Abdulrahman University, P.O. Box 84428, Riyadh 11671, Saudi Arabia (PNU) and Department of Mathematics and Statistics, University of Strathclyde, Glasgow G1 1XH, United Kingdom
}
\author{Apala Majumdar}
\email{apala.majumdar@manchester.ac.uk}
\affiliation{Department of Mathematics, University of Manchester, Manchester M13 9PL, United Kingdom}

\begin{abstract}
We study bent-core nematic (BCN) systems in two-dimensional (2D) and three-dimensional (3D) settings, focusing on the role of cybotactic clusters, phase transitions, confinement effects and applied external fields. We propose a generalised version of Madhusudana's two-state model for BCNs in \cite{madhusudana2017two} with two order parameters: $\Q_g$ to describe the ambient ground-state (GS) molecules and $\Q_c$ to describe the additional ordering induced by the cybotactic clusters. The equilibria are modelled by minimisers of an appropriately defined free energy, with an empirical coupling term between $\Q_g$ and $\Q_c$. We demonstrate two phase transitions in spatially homogeneous 3D BCN systems at fixed temperatures: a first-order nematic-paranematic transition followed by a paranematic-isotropic phase transition driven by the GS-cluster coupling. We also numerically compute and give heuristic insights into solution landscapes of confined BCN systems on 2D square domains, tailored by the GS-cluster coupling, temperature and external fields. This benchmark example illustrates the potential of this generalised model to capture tunable director profiles, cluster properties and potential biaxiality induced by antagonistic $\Q_g$ and $\Q_c$-profiles.
\end{abstract}

\maketitle

\section{Introduction} \label{sec:intro}

Liquid crystals (LCs) are fascinating mesophases with physical properties intermediate between the conventional solid and liquid phases \cite{degennes}. LCs are partially ordered materials, i.e., they are more ordered than liquids and less ordered than crystalline solids. The partial ordering can manifest in special material directions, layered structures, chirality, columnar structures etc. Nematic liquid crystals (NLCs) are the simplest and perhaps, the most common type of LCs composed of rod-like molecules that tend to preferentially align along some distinguished directions, referred to as \emph{directors}, rendering long-range orientational order \cite{degennes}. Smectic LCs are effectively layered LCs, i.e., the LC molecules organise into layers and there is a preferred \emph{director} or nematic ordering within each layer. Then, we have cholesterics which are twisted nematics such that the director naturally twists in space and imparts macroscopic chirality. Molecular shape can have strong consequences on macroscopic LC properties and recently, there has been tremendous interest in bent-core nematics (BCNs) \cite{jaklirevmodernphysics2018}. In non-technical terms, a BCN molecule is composed of two connected rods with an opening angle. This kinked shape opens a plethora of possibilities - new phases, biaxiality, natural chirality, non-Newtonian rheology and exotic morphologies \cite{jaklirevmodernphysics2018, patranabish2019ref1, patranabish2019ref2}. 

BCNs were discovered in the 1990's; they are remarkable materials that can exhibit different types of order --- polar, chiral, octupolar, biaxial order etc. They can exhibit structural transitions to more ordered phases —such as twist‑bend and smectic—depending on temperature and molecular architecture 
~\cite{jaklirevmodernphysics2018}. These BCN phases offer tremendous potential for electro-optic devices, rheological applications and novel functional materials. Cybotactic clusters are an intrinsic feature of BCNs; they are smectic-like nanoclusters such that BCN molecules get locked into clusters because of their kinked shape. The existence of cybotactic clusters has been confirmed by multiple experimental methods, e.g., SAXS, cryo-TEM methods, electro-optic measurements  \cite{jaklipage102009, jaklipage10shankar, jaklipage10gao} etc. Cybotactic clusters are small, usually tens of nanometers in lateral dimensions and persist in the nematic and sometimes, even in the isotropic BCN phase. Cybotactic clusters render additional ordering to the conventional nematic phase and this ordering clearly manifests in optical images (four-lobed structures, characteristic of local smectic‑C‑like ordering, in SAXS measurements), rheological properties (non-Newtonian properties), elastic properties (suppressed twist elastic constants leading to spontaneous chirality), macroscopic biaxiality \cite{jaklirevmodernphysics2018} etc.  As such, it is essential to quantify the additional ordering induced by the cybotactic clusters and how this ordering modulates the conventional ambient nematic ordering.

In \cite{madhusudana2017two}, Madhusudana proposes a two-state model for BCNs, based on the hypothesis that there are two types of molecules: the ground-state (GS) molecules that constitute the out-of-cluster ambient environment and the excited molecules (ES) that constitute the cybotactic clusters. There are two scalar order parameters, $S_g$ and $S_c$, to describe the orientational ordering of the GS molecules and the ES molecules respectively. Madhusudana assumes spatially uniform $S_g$ and $S_c$ profiles, i.e., he works with spatially homogeneous BCN systems along with various other assumptions --- fixed time-independent and temperature-independent cluster sizes, an empirical coupling term between $S_g$ and $S_c$ and no explicit information about the BCN directors. Madhusudana proposes a phenomenological free energy comprising two Ginzburg-Landau type potentials for $S_g$ and $S_c$ with various temperature-dependent and material-dependent parameters and an empirical GS-ES coupling term that is parameterised by a coupling coefficient. The physically observable spatially homogeneous BCN configurations are modelled by minimisers of this two-state free energy. The most important finding concerns the existence of a new paranematic phase with small values of $S_g$ and $S_c$, just before the transition to the disordered isotropic phase.  This model also effectively captures other experimental trends reported in \cite{wiantetal2008} --- enhanced order parameters induced by the cybotactic clusters, a nematic-paranematic phase transition followed by a paranematic-isotropic phase transition with increasing temperature and non-linear dependence of order parameters on external magnetic fields.

However, Madhusudana's model cannot account for confinement effects or spatial inhomogeneities. In \cite{patranabish2019one}, the authors propose a one-dimensional model for spatially inhomogeneous BCN samples, including elastic energy terms for $S_g$ and $S_c$. They study BCN samples in channel geometries with fixed boundary conditions for $S_g$ and $S_c$; their numerical results demonstrate that confinement or geometric frustration promotes interior clustering or enhances interior ordering and that the GS-ES coupling can promote ordered director profiles even at high temperatures contrary to the expectation of isotropy or disorder at high temperatures. In \cite{mondal2025}, the authors study the effects of doping on cybotactic clusters along with the emergence of polar order in such BCN systems. They perform extensive experiments on phase transitions, birefringence, order parameter measurements and dielectric spectroscopy data and propose a simple Madhusudana-type free energy to partially describe their experimental results. 
In this paper, we build on these works to propose a generalised free energy for confined BCN systems in two-dimensional (2D) and three-dimensional (3D) settings. By analogy with Landau-de Gennes theory, we define two tensor order parameters: $\Q_g$ to describe the orientational ordering of the GS molecules and $\Q_c$ to describe the additional ordering induced by the ES molecules within cybotactic clusters \cite{majumdarejam2010}. The key advantage is that $\Q_g$ and $\Q_c$ not only contain information about $S_g$ and $S_c$ but also contain information about the GS and ES directors, that are the preferred directions of average alignment of the GS molecules and new distinguished material directions induced by the cybotactic clusters. The free energy has an elastic energy to penalise spatial inhomogeneities and a bulk energy that determines the GS and ES ordering as a function of temperature, GS-ES coupling and other parameters, along with terms that account for external (electric or magnetic) fields. The bulk energy is a direct generalisation of Madhusudana's free energy and is composed of Ginzburg-Landau type potentials for $\Q_g$ and $\Q_c$ respectively and a $\Q_g$-$\Q_c$ coupling term. The Ginzburg-Landau potentials dictate that $S_g$ and $S_c$  increase with decreasing temperatures and approach zero with increasing temperature (the isotropic or disordered phase). The GS-ES coupling and external field terms can strongly boost ordering, even for high temperatures. The physically observable textures are modelled by minimisers of this generalised energy, subject to the imposed boundary conditions. The minimisers and the critical points of the free energy are solutions of boundary-value problems for the associated Euler-Lagrange equations --- typically a system of coupled and nonlinear partial differential equations and the rest of this manuscript is devoted to the study of critical points of this generalised free energy in physically motivated settings.

The rest of the paper is organised as follows. In Section~\ref{sec:Ldg}, we introduce a generalised Landau-de Gennes type free energy for BCN systems. In Section~\ref{sec:phasetransitions}, we study phase transitions in spatially homogeneous BCN samples, with and without external fields, in 2D and 3D. Notably, we find two phase transitions in spatially homogeneous 3D BCN samples driven by increasing GS-ES coupling strength, at high temperatures. The GS-ES coupling strength is measured in terms of a parameter $\gamma$. If $\gamma=0$, then $S_g = S_c=0$ for spatially homogeneous samples at high temperatures (defined by $A>0$ in (\ref{eq:F})). As $\gamma$ increases, the energy minimisers have positive values of $S_g$ and $S_c$ even for high temperatures. At a fixed high temperature, increasing $\gamma$ has a strong ordering effect that counteracts the disordering effects of high temperature. This is analogous to the nematic-paranematic phase transition and the paranematic-isotropic phase transition with increasing temperature, at fixed $\gamma$, as reported in \cite{madhusudana2017two}. The external fields typically boost the values of $S_g$ and $S_c$, shift the bifurcation points and we lose the paranematic-isotropic phase transition. Given that the values of $S_g$ and $S_c$ are very small in the paranematic phase, it is unclear if the paranematic phase would be distinguishable from the conventional isotropic phase in experiments. In Section~\ref{sc:square}, we study solution landscapes of confined BCN systems on a square domain with Dirichlet boundary conditions for $\Q_g$ and $\Q_c$. This is a benchmark  well-studied example for conventional nematics; see the papers \cite{yin2020construction, luomultistability2012, kusumaatmaja2015}. A solution landscape is a network of critical points of the free‑energy surface, revealing possible pathways between the different stable solutions and unstable saddle points of the free energy. The unstable saddle points play a crucial role in the system dynamics. We study the interplay between domain size, temperature, the GS-ES coupling and external field strength on the energy minimisers and the associated solution landscapes.
We recover some familiar $\Q_g$ profiles from the studies on conventional nematics, some new $\Q_g$ profiles tailored by the additional physics in this model, co-aligned $(\Q_g, \Q_c)$ profiles with co-aligned GS and ES directors and antagonistic $(\Q_g, \Q_c)$-profiles with mutually perpendicular GS and ES directors. Our simple examples illustrate the tremendous potential of tuning model parameters to get exotic morphologies, some of which could lead to macroscopic biaxiality e.g., the antagonistic $(\Q_g, \Q_c)$ profiles. We conclude with a discussion of the limitations of the model and future improvements in Section~\ref{sec:conclusions}.

\section{Landau-de Gennes Modelling for Bent-Core Nematic Systems}
\label{sec:Ldg}
In this paper, we propose a two-state generalised Landau-de Gennes (LdG) type free energy for confined BCN systems building on Madhusudana's model in \cite{madhusudana2017two}. We define a GS LdG order parameter, $\mathbf{Q}_g$, which is a symmetric, traceless $d\times d$ matrix, where $d$ is the spatial dimension. The eigenvectors of $\mathbf{Q}_g$ model the GS nematic directors, interpreted as the locally preferred directions of averaged GS molecular alignment in space and the associated eigenvalues measure the degree of orientational ordering about the GS directors. We define the GS director to the eigenvector of $\mathbf{Q}_g$ with the largest positive eigenvalue. 
The GS is said to be in the isotropic phase if $\Q_g = 0$. It is in the uniaxial phase if $\Q_g$ has two equal non-zero eigenvalues for $d=3$ and $\Q_g$ can be written as
\begin{equation}
\Q_g = \sqrt{\frac{3}{2}}S_g\left(\n_g\otimes \n_g - \frac{I_3}{3}\right)\end{equation}
where $\n_g$ is the eigenvector with the non-degenerate eigenvalue and $S_g$ is proportional to the eigenvalue associated with $\n_g$. If $S_g>0$, then $\n_g$ is the GS director. $\Q_g$ can be biaxial for $d=3$ (and not for $d=2$) when it has three distinct eigenvalues. 
In 2D,
\begin{gather}
 \ \Q_g = \sqrt{2} S_g \left(\n_g\otimes \n_g - \frac{I_2}{2} \right),\label{eq:ng_2D}
\end{gather} 
where $\n_g$ is the GS director and $S_g$ is positive. When $d=2$, the defect set is identified with the nodal set of $S_g$. For completeness, $I_d$ is the identity matrix in $d$ dimensions.

We treat the cybotactic clusters as uniformly distributed sub-nanometric/very small clusters, such that the size of the cluster is much smaller than the spacing between the clusters and the volume fraction of the clusters is much smaller than unity. This is the so-called ``dilute limit" in homogenization theory \cite{canevarihomogenisation}, consistent with the assumptions in Madhusudana's paper. In this case, the highly localised cybotactic clusters interact with the surrounding GS molecules by means of an averaged field, the $\mathbf{Q}_c$ - tensor field, referred to as the cluster order parameter in this paper. In principle, $\mathbf{Q}_c$ contains information about the cluster shapes and the GS-ES interactions, assuming that the clusters do not interact with each other. 
For example, in $d=2$, we have
\begin{gather}
 \ \Q_c = \sqrt{2} S_c \left(\n_c\otimes \n_c - \frac{I_2}{2} \right),\label{eq:nc_2D}
\end{gather} where $\n_c$ is referred to as the \emph{ES/cluster director} and $S_c$ is positive. As an immediate consequence, we have two tensor order parameters, $\mathbf{Q}_g$ and $\mathbf{Q}_c$, as opposed to two scalar order parameters, $S_g$ and $S_c$ in \cite{madhusudana2017two}, that can describe complex information about the GS and cluster directors, structural transitions and confinement effects in inhomogeneous systems which are outside the scope of the two-state model in \cite{madhusudana2017two}.


The corresponding generalised LdG-type free energy is
\begin{equation}\label{GM}
    \begin{aligned}
       &F(\Q_g,\Q_c)  = \int_{\Omega }K_g |\nabla \Q_g |^2  + K_c |\nabla \Q_c|^2 + (1-a_x)\\
       & \times \left\{ \frac{a_g}{2}(T - T^*)\tr\Q_{g}^2 - \frac{B_g }{3}\tr\Q_{g}^3 + \frac{C_g}{4}(\tr\Q_{g}^2)^2\right.\\
       &\left.-\sqrt{2}E_{el}\e^T\Q_{g}\e \right\} \\
       & + \frac{a_x}{N_c}\left\{-(1-a_x)\gamma \tr(\Q_g \Q_c) + \frac{\alpha_c}{2}\tr\Q_{c}^2 + \frac{\beta_c}{4} (\tr\Q_{c}^2)^2 \right\}\\
       &- \sqrt{2}a_x J E_{el}\e^T\Q_c\e ~ d\x,
    \end{aligned}
\end{equation} where $\Omega$ is a $d$-dimensional bounded domain with Lipschitz boundary and $d=2,3$. The first two terms are the elastic energy density terms associated with $\mathbf{Q}_g$ and $\mathbf{Q}_c$ respectively, i.e., we assume a one-constant elastic energy density to penalise spatial inhomogeneities in $\mathbf{Q}_g$ and $\mathbf{Q}_c$ such as strong GS-ES interactions, interior defects, geometric frustration etc. $K_g$ and $K_c$ denote the positive elastic constants associated with $\mathbf{Q}_g$ and $\Q_c$ respectively. We note that other choices of anisotropic elastic energy densities are also possible, see \cite{han2022elastic}. For BCN materials, the bend elastic constant 
$K_{33}$ is often considerably smaller than the splay ($K_{11}$) or twist ($K_{22}$) constants, reflecting an intrinsic bend‑softening associated with the molecular geometry. The mole fraction of the ES molecules $a_x$ is given by
\begin{equation*}
    a_x = \exp(-E_{ex}/k_B T)\Bigg/\left[1+\exp\left(\frac{-E_{ex}}{k_B T}\right)\right],
\end{equation*}
as $E_{ex}$ is the excitation energy of the ES molecules,
$k_B$ is Boltzmann’s constant and $T$ is the temperature. $a_x = 0.1$ throughout the paper consistent with \cite{patranabish2019one}. 
Further, there maybe an intrinsic $a_x$-dependence of the elastic constants. In the interest of brevity and accounting for uncertainties in the properties of the cybotactic clusters, including their elastic constants, we use the one-constant elastic energy density as a first approximation. 
 The parameters $a_g$, $B_g$, $C_g$, and $T^*$ are the usual LdG parameters that describe the first-order nematic-isotropic transition for the GS molecules and $\gamma$ is the coupling parameter between the GS molecules and the clusters \cite{patranabish2019one,madhusudana2017two}. We note that the term $\textrm{tr}\left(\mathbf{Q}_g \mathbf{Q}_c \right)$ is a generic coupling term for systems with two order parameters that is consistent with frame-indifference and material symmetry requirements \cite{canevarihomogenisation}. 
$\alpha_c$ and $\beta_c$ are coefficients for the saturation terms to ensure that $|\mathbf{Q}_c|^ 2$ is reasonably bounded, for at least a certain range of $\gamma$. There is no evidence of first-order isotropic-nematic phase transitions within cybotactic clusters to date, so the term $tr\Q_c^3$ is absent. We also note that Madhusudana's original work does not include a cubic term in $S_c$, analogous to a $tr \Q^3_c$ term. In the subsequent numerical simulations, we adjust the values of $\alpha_c$ and $\beta_c$ so that the resulting values of $S_c$ are reasonable and not too large.
$N_c$ is the number of ES molecules within each cluster.
We also consider the effects of an external field, $\mathbf{E} = E_{a} \e$, whose direction is modelled by a unit-vector
$\e$, and the magnitude is $E_a$. $E_{el}$ is the external field energy density $( \frac{1}{2}\epsilon_0 \Delta \epsilon E_a^2)$ where $\epsilon_0$ is the free-space permittivity, $\Delta \epsilon$ is a material anisotropy parameter (dielectric or magnetic) for GS molecules and $J$ accounts for the larger shape anisotropy of ES molecules, i.e., $J>1$. In particular, the GS molecules tend to align with $\mathbf{E}$ if $\Delta \epsilon >0$ and orthogonal to $\mathbf{E}$ if $\Delta\epsilon <0$ \cite{degennes}.

We nondimensionalise the free energy above with $\tilde{x} = x/\lambda$ and assume $K_g =K_c$, where $\lambda$ is a characteristic length scale of the $d$-dimensional domain $\Omega$,
\begin{equation}
\begin{aligned}
   &\tilde{F}(\Q_g,\Q_c) = \frac{\lambda^{2-d}F}{K_g}= \int_{\tilde{\Omega}} |\tilde{\nabla} \Q_g |^2  + |\tilde{\nabla} \Q_c|^2 + 2\tilde{\lambda}^2 f_b d\tilde{\x}. \label{eq:F}
\end{aligned}
\end{equation}
where
\begin{equation}
\begin{aligned}
f_b &= \frac{A}{2}\tr\Q_{g}^2 - \frac{B}{3}\tr\Q_{g}^3 + \frac{C}{4}(\tr\Q_{g}^2)^2- \sqrt{2}E\e^T\Q_g \e \\
   &+ \frac{M}{2}\tr\Q_{c}^2 + \frac{N}{4} (\tr\Q_{c}^2)^2 -D \tr(\Q_g \Q_c) - \sqrt{2}PE\e^T\Q_c \e 
\end{aligned}
\end{equation}
where
\begin{gather}
A = (1-a_x)a_g(T-T^*)/C_g, B = (1-a_x)B_g/C_g, \nonumber\\
C = (1-a_x), D = \frac{a_x}{N_c}(1-a_x)\gamma/C_g, E= (1-a_x)E_{el}/C_g, \nonumber\\
M = \frac{a_x}{N_c}\alpha_c/C_g, N = \frac{a_x}{N_c}\beta_c/C_g, P = a_x J/(1-a_x), \nonumber\\
\tilde{\lambda}^2 = C_g\lambda^2/(2K_g),\nonumber
\end{gather} and $\tilde{\Omega}$ is the rescaled domain.
For comparison with previous work on one-dimensional BCN systems in \cite{patranabish2019one}, we set $A = \pm0.04$, $B = 0.34$, $C = 0.9$, $D = 2\times 10^{-3}$, $E = 1.6\times 10^{-4}$,   $M = 9.7\times 10^{-5}$, $N = 1.78\times 10^{-3}$, $P = 0.133$ which are derived from
$K_g =K_c = K = 15\mathrm{pN} = 15 \times 10^{-12}\mathrm{N} = 15 \times 10^{-7} \mathrm{dyn} \text{(under one constant approximation)}$;
$a_g = 0.04 \times 10^7/4 \ \mathrm{ergs/(cm^3K)} $,
$B_g = 1.7 \times 10^7/4 \ \mathrm{ergs/cm^3}$,
$C_g =4.5 \times 10^7/4 \ \mathrm{ergs/cm^3}$,
$\alpha_c = 0.22 \times 10^7/4 \ \mathrm{ergs/cm^3}$,
$\beta_c = 4.0 \times 10^7/4 \ \mathrm{ergs/cm^3}$,
and $\gamma = 5.0 \times 10^7/4 \ \mathrm{ergs/cm^3}$;
$T^*  = 355 K$,
$N_c = 50$,
$J =1.2$,
$E_{el} = 2000 \mathrm{ergs/cm^3}$,
$E_{ex} = 1.1 \times 10^{-13} \mathrm{ergs}$,
and $T =360\mathrm{K}$ (for the high temperature case with $ A > 0$ ) and $350 \mathrm{K}$ (for the low temperature case with $A < 0$). 
In subsequent sections, we focus on phase transitions in spatially homogeneous systems and confinement effects in square domains as a function of temperature $A$, the GS-ES coupling strength measured by $D$ and external field strength, $E$.

For clarity, we temporarily refer to the 
$F$ from Eq. \eqref{GM}, where $\Omega$ is a $d$-dimensional region, as $F_d$.
In three dimensions, the free energy $F_3$ has the unit of energy (erg),
whereas in two dimensions, $F_2$ has the unit of force (N). 
This effectively assumes a uniform extension along the third dimension, 
so the true three‑dimensional energy can be written as 
$F_3 = F_2 \times (\text{thickness})$. 
After nondimensionalization in Eq.~\eqref{eq:F}, however, the factor 
$\lambda^{2-d}$, which depends on the spatial dimension $d$, 
renders the dimensionless free energy $\tilde{F}$ unit‑free 
in both two and three dimensions.

\section{Phase Transitions in the Generalised LdG model in 3D and 2D settings}
\label{sec:phasetransitions}
In \cite{madhusudana2017two}, the author studies phase transitions in spatially homogeneous BCN systems as a function of temperature, based on the two-state free energy in terms of $S_g$ and $S_c$. 
The generalised LdG energy in \eqref{GM} reduces to  Madhusudana's two-state model if we assume that $\Q_g$ and $\Q_c$ are spatially constant uniaxial tensors and the directors $\n_g = \n_c = \e \equiv constant$, i.e., the bulk energy density in \eqref{eq:F} reduces to
\begin{equation}
\begin{aligned}
f_{b} &= \frac{A}{2}S_g^2 - \frac{B}{3}S_g^3 + \frac{C}{4}S_g^4  - ES_g\\
&+ \frac{M}{2}S_c^2 + \frac{N}{4}S_c^4 - DS_gS_c - PE S_c, \label{eq:fb}
\end{aligned}
\end{equation}
where $B = 0$ in 2D settings. Equation \eqref{eq:fb} is the two-state free energy studied in \cite{madhusudana2017two}.
The critical points of this bulk energy density are the solutions, $(S_g, S_c)$, of the following system of bivariate polynomial equations:
\begin{align}
AS_g - BS_g^2+CS_g^3-DS_c-E=0,\label{eq:Sg_Sc1}\\
MS_c + NS_c^3-DS_g-PE=0.\label{eq:Sg_Sc2}
\end{align}
We use an open-source computational knowledge engine, WolframAlpha \cite{Wolfram}, to obtain the numerical solutions for the nonlinear system of equations \eqref{eq:Sg_Sc1}-\eqref{eq:Sg_Sc2}. Since \eqref{eq:Sg_Sc2} yields $S_g$ as a cubic function of $S_c$, we substitute it into \eqref{eq:Sg_Sc1} to obtain a ninth-degree polynomial in $S_c$. This results in nine solutions for $S_c$. Here we only consider the real ones.
The stability of the critical points is determined by the eigenvalues of the Hessian matrix 
\begin{equation*}
    \begin{pmatrix}
A - 2BS_g + 3C S_g^{2} & -D \\
-D & M + 3 N S_c^{2}
\end{pmatrix}
\end{equation*}
If both eigenvalues are positive, the critical point is stable; if there is at least one negative eigenvalue, the critical point is unstable.

\begin{figure*}
    \centering
    \includegraphics[width=0.9\textwidth]{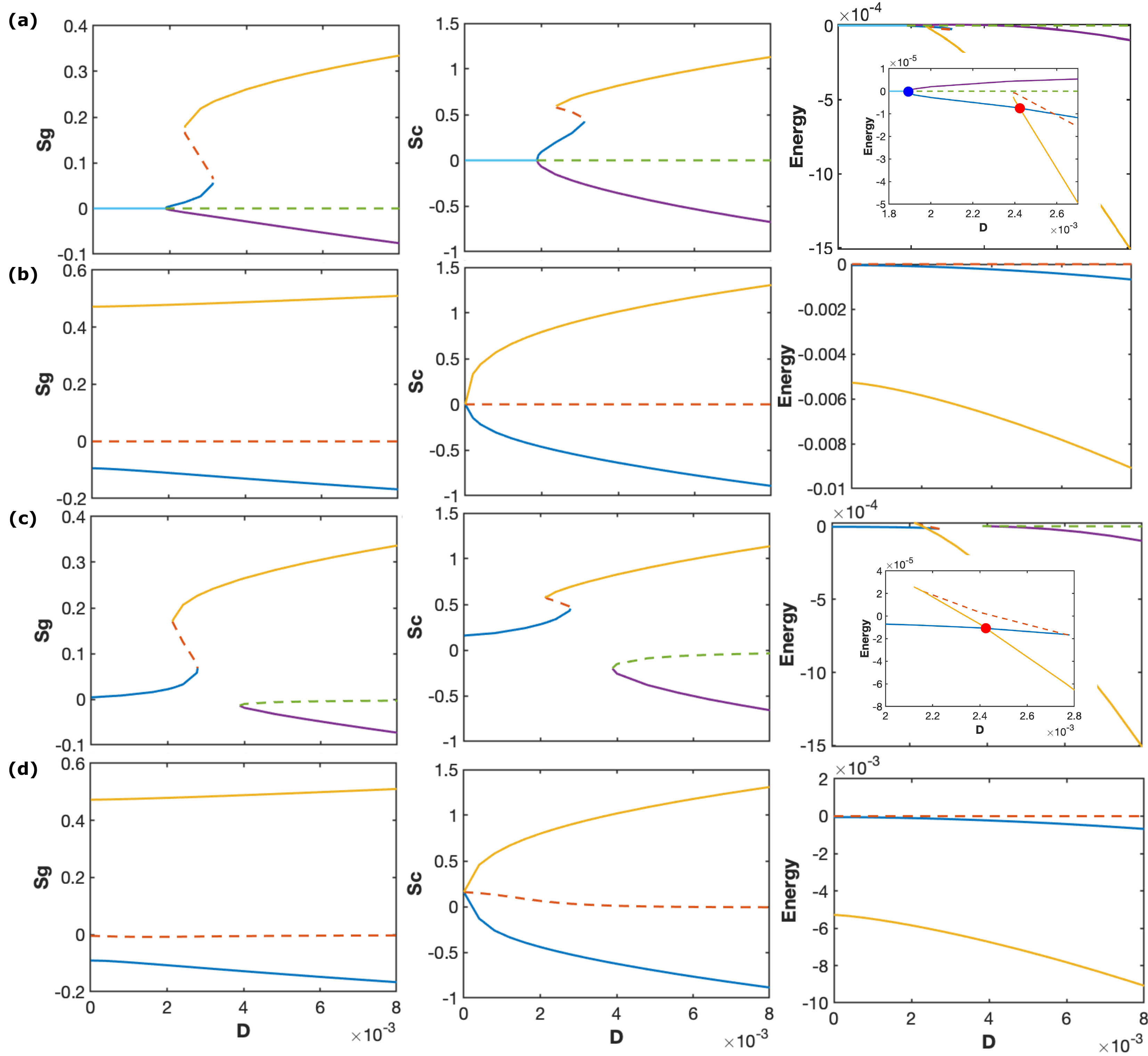}
    \caption{$S_g$, $S_c$, and energy plots of solutions for \eqref{eq:Sg_Sc1}-\eqref{eq:Sg_Sc2} vs $D\in[0,8\times 10^{-3}]$ corresponding to $\gamma\in[0,5\times 10^8] \mathrm{ergs/cm^3}$ in 3D setting ($B = 0.34$). In (a-b), a zero external field with $E=0$ is used, while in (c-d), a nonzero external field with $E=1.6\times 10^{-4}$ is applied. In (a, c), $A = 0.04$ is fixed whereas in (b, d), $A = -0.04$ is fixed. The dashed lines represent unstable solutions while the solid lines represent stable solutions. The red dots indicate the first‑order phase transitions in (a) and (c). The blue dot indicates the second‑order phase transition in (a).
    }
  \label{fg:Fig1}
\end{figure*}

In the 3D setting, with a zero external field ($E=0$), the isotropic phase $S_g = S_c =0$ is always a solution of the coupled algebraic equations \eqref{eq:Sg_Sc1}-\eqref{eq:Sg_Sc2}, for any reduced temperature $A$ (Fig. \ref{fg:Fig1}(a-b)).
At the characteristic high temperature $A = 0.04$ (Fig. \ref{fg:Fig1}(a)),  the isotropic solution, $(S_g, S_c) = (0,0)$, is the stable solution with the lowest energy for $D$ sufficiently small. When $D\approx 1.88\times 10^{-3}$, the isotropic solution loses stability and undergoes a pitchfork bifurcation into two stable solutions: a weakly ordered solution \cite{madhusudana2017two} with small $S_g>0$ and $S_c>0$ (solid blue line) and a negative solution branch with $S_g<0$ and $S_c<0$ (solid purple line). We interpret this weakly ordered stable solution as the \emph{paranematic-type phase}.
A bifurcation point is a point at which a small, smooth change in a parameter value causes a sudden qualitative change in the behaviour of the system, such as the appearance or disappearance of critical points and changes in their stability.

As $D$ further increases, we observe a saddle-node bifurcation accompanied by the creation of a stable nematic solution with large $S_g>0$ and $S_c>0$ (solid yellow line) and one unstable solution (dashed orange line). When $D\approx 3\times 10^{-3}$, the orange unstable branch and blue \emph{paranematic} stable branch merge and disappear. Then the number of solutions changes from five to three for $D$ large enough. According to the energy plot in Fig. \ref{fg:Fig1}(a), there are two phase transitions as $D$ increases. The phase transition occurs at a value of $D$ for which there are two solution branches with equal free energies in \eqref{eq:fb}. When $D\approx 1.88\times 10^{-3}$, there is a second-order phase transition between the isotropic phase and the weakly ordered \emph{paranematic-type} phase. When $D\approx 2.4\times 10^{-3}$, there is a first-order phase transition between the weakly ordered \emph{paranematic-type} phase and ordered nematic phase.
For a first-order phase transition, the order parameters $S_g$ or $S_c$ are discontinuous.
For a second-order phase transition, $S_g$ and $S_c$ are continuous, while the derivatives of $S_g$ or $S_c$ with respect to $D$ are discontinuous. If we informally interpret $D$ as promoting order or as being an analogue of low temperatures, then this is consistent with the picture in \cite{madhusudana2017two}  --- a first-order nematic-paranematic transition with increasing temperature (decreasing $D$) followed by the conventional isotropic-(para)nematic phase transition with increasing temperature (decreasing $D$). We note that $S_c$ is significantly larger than $S_g$ in magnitude for all solution branches.

At the characteristic low temperature $A = -0.04$ (Fig. \ref{fg:Fig1}(b)), we always have three solution branches --- the unstable isotropic solution with $S_g=S_c=0$ (dashed line in orange), positive stable solution branch with positive $S_g$ and $S_c$ (solid yellow line) and negative stable solution branch with negative $S_g$ and $S_c$ (solid blue line). The positive solution branch always has the lowest energy in the solution set, corresponding to well-oriented configurations consistently observed in experiments. We note that $S_g<0$ (or $S_c<0$) describes a state where the GS (ES) molecules are relatively randomly oriented in the plane perpendicular to $\n_g$ ($\n_c$), where $\n_g$ is the eigenvector of $\Q_g$ with the negative non-degenerate eigenvalue.

When an external field is applied ($E = 1.6 \times 10^{-4}$), the isotropic solution branch no longer exists (Fig. \ref{fg:Fig1}(c-d)). For $A=0.04$ (Fig. \ref{fg:Fig1}(c)), the bifurcation structure and the number of solutions change compared to the zero external field (with $E=0$). When $D$ is small enough, there is a unique weakly-ordered \emph{paranematic}-type solution with positive $S_g$ and $S_c$ close to zero. When $D\approx 2.1\times 10^{-3}$, a stable solution branch with relatively large positive values of $S_g$ and $S_c$, and an unstable solution branch appear through a saddle-node bifurcation. When $D\approx 2.8\times 10^{-3}$, the stable weakly-ordered \emph{paranematic} solution and the unstable solution merge and disappear. When $D\approx 4\times 10^{-3}$, we note the simultaneous appearance of a stable and unstable negative solution branch, both of which have $S_g<0$ and $S_c<0$. According to the energy plot in Fig. \ref{fg:Fig1}(c),  there is only one first-order phase transition between the weakly-ordered \emph{paranematic} phase (solid blue line) and strongly ordered nematic solution (solid yellow line). Again, decreasing $D$ is analogous to increasing temperature in \cite{madhusudana2017two}.

For the characteristic low temperature $A = -0.04$ (Fig. \ref{fg:Fig1}(d)), we still retain three solution branches and the global bulk energy minimum is attained by the positive solution branch with positive $S_g$ and positive $S_c$. We note that $S_g$ and $S_c$ increase with increasing $E$, along the solid yellow solution branch in Fig. \ref{fg:Fig1}(b) and (d).

\begin{figure*}
    \centering
    \includegraphics[width=0.9\textwidth]{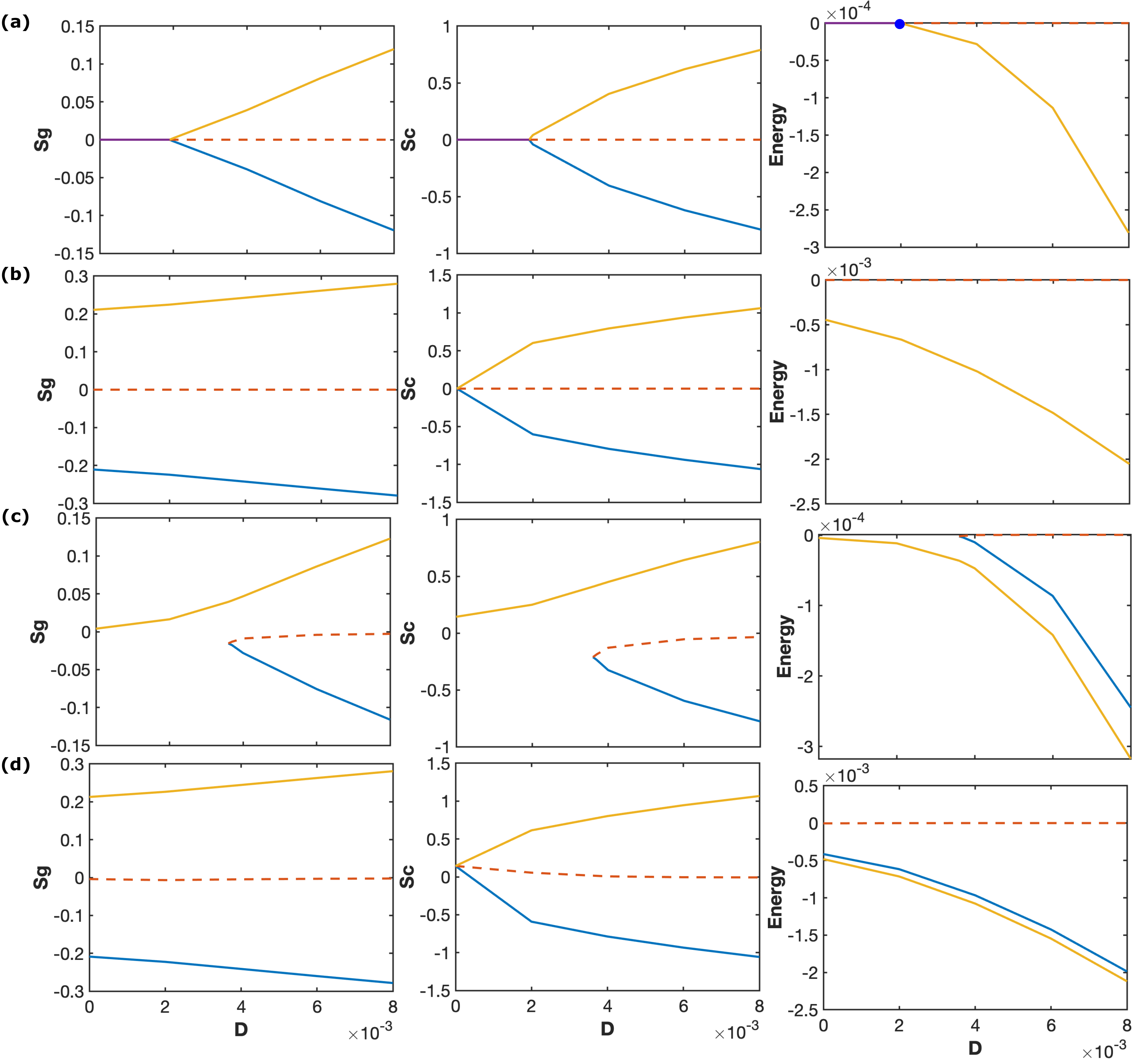}
    \caption{$S_g$, $S_c$ and energy plots of solutions for \eqref{eq:Sg_Sc1}-\eqref{eq:Sg_Sc2} vs $D\in[0,8\times 10^{-3}]$ corresponding to $\gamma\in[0,5\times 10^8] \mathrm{ergs/cm^3}$ in 2D setting ($B = 0$). In (a-b), a zero external field with $E=0$ is used while in (c-d), a nonzero external field with $E=1.6\times 10^{-4}$ is applied. In (a, c), $A = 0.04$ is fixed whereas in (b, d), we fix $A = -0.04$. The dashed lines represent unstable solutions while the solid lines represent stable solutions. The blue dot indicates the second‑order phase transition in (a).
    }
  \label{fg:Fig2}
\end{figure*}
In the 2D setting in Fig. \ref{fg:Fig2}, we set $B=0$ in the algebraic equations \eqref{eq:Sg_Sc1}-\eqref{eq:Sg_Sc2}. Compared to the 3D setting in Fig. \ref{fg:Fig1}, we lose the weakly-ordered paranematic phase and the associated saddle-node bifurcations. 
Without an external field ($E=0$), there are
three solution branches --- the isotropic solution branch with $S_g = S_c= 0$; the positive nematic solution branch with positive $S_g$ and $S_c$, and the negative solution branch with $S_g$ and $S_c <0$. Since $B = 0$ and $E = 0$, the positive and negative solution branches have equal free energies for $A = \pm 0.04$. For $A = 0.04$, the isotropic solution is stable when $D$ is small enough. When $D$ increases further, the isotropic solution loses stability and bifurcates into stable positive and negative ordered solutions. Compared to the 3D setting in Fig. \ref{fg:Fig1}(a), we only retain the second-order phase transition in Fig. \ref{fg:Fig2}(a). There is no bifurcation for $A = -0.04$ and the isotropic solution branch is unstable for all values of $D$ under consideration (Fig. \ref{fg:Fig2}(b)).

With a nonzero external field ($E=1.6 \times 10^{-4}$), the positive nematic solution branch is energetically preferred for high and low temperatures. The positive solution branch corresponds to the GS and ES molecules being aligned with the external field. The negative solution branch (with $S_g, S_c <0$) corresponds to the GS and ES molecules being oriented orthogonal to the external field. For $A = 0.04$ (Fig. \ref{fg:Fig2}(c)), there is no phase transition in the 2D setting, contrary to the 3D setting which exhibits a first-order paranematic-nematic phase transition with increasing $D$. The globally stable solution branch is always the ordered nematic phase (solid yellow line). There are no bifurcations for $A = -0.04$, analogous to the $E=0$ case but the symmetry between the positive and negative solution branches is broken in the energy plot in Fig. \ref{fg:Fig2}(d).

We conclude this section by demonstrating that the generalised bulk energy in \eqref{GM} can only have isotropic or uniaxial critical points for $d=3$, i.e., bulk biaxiality is outside the scope of this relatively simple generalised bulk energy. The critical points of the generalised bulk energy density (with $E=0$)
\begin{multline}
    f_b = \frac{A}{2}\tr\Q_{g}^2 - \frac{B}{3}\tr\Q_{g}^3 + \frac{C}{4}(\tr\Q_{g}^2)^2 \\
   + \frac{M}{2}\tr\Q_{c}^2 + \frac{N}{4} (\tr\Q_{c}^2)^2 -D \tr(\Q_g \Q_c)  
\end{multline}
are solutions of the coupled systems of partial differential equations:
\begin{align}
&A \Q_g - B\left( \Q_g^2 - \frac{1}{3}\tr\Q_g^2 \mathbf{I} \right) + C \left(\tr\Q_g^2\right) \Q_g = D \Q_c,\label{eq:Qg}\\
&\left(M  + N \tr\Q_c^2\right) \Q_c = D \Q_g.\label{eq:Qc}
\end{align}
Setting $\sigma = (M + N\tr\Q_c^2)/D$ and substituting Eq. \eqref{eq:Qc} into Eq. \eqref{eq:Qg} yields
\[
A \sigma \Q_c - B \sigma^2 \left( \Q_c^2 - \frac{1}{3}\tr\Q_c^2\mathbf{I} \right) + C \sigma^3 \left(\tr\Q_c^2 \right)\Q_c = D \Q_c.
\]
Let $\Q_c$ be a diagonal $3\times 3$ matrix with diagonal entries, $Q_1, Q_2, Q_3$ respectively and zero non-diagonal entries. The above equation holds for all matrix components of $\Q_c$ and repeating the arguments from \cite{majumdarejam2010}, we deduce that there are at least two equal diagonal entries. This implies that $\Q_c$ has at least two equal eigenvalues and hence all critical points of the generalised bulk energy density are either isotropic or uniaxial pairs, $(\Q_g, \Q_c)$, as outlined above.
Hence, the generalised bulk energy density does not admit biaxial critical points in $d=3$ and bulk biaxiality is outside the scope of this model.

\section{Confinement Effects for BCN Spatial Equilibria on Square Domains }\label{sc:square}
Next we consider a thin three-dimensional square well filled with a prototype BCN material:
\[
\Omega \times [0,h]
\] where $\Omega$ is a square domain and $h$ is the height of the well.
In the $h\to 0$ limit and for certain choices of the surface energies, one can prove that conventional LdG energy minimisers have a fixed eigenvector in $\mathbf{z}$-direction with a fixed constant eigenvalue and all dependent variables are independent of the $\mathbf{z}$-coordinate \cite{golovaty2015dimension}, so that it suffices to consider the reduced LdG tensor
 --- a symmetric, traceless $2\times 2$ matrix with only two degrees of freedom. In other words, the LdG $\Q$-tensor can be written as
\begin{equation*}
\begin{aligned}
\Q &= s_1(\mathbf{x}\otimes\mathbf{x}-\mathbf{y}\otimes\mathbf{y}) + s
_2(\mathbf{x}\otimes\mathbf{y}+\mathbf{y}\otimes\mathbf{x})\\
&+s_3(2\mathbf{z}\otimes\mathbf{z}-\mathbf{x}\otimes\mathbf{x} - \mathbf{y}\otimes\mathbf{y}).
\end{aligned}
\end{equation*}
where $\mathbf{x} = (1,0,0)^T$, $\mathbf{y} = (0,1,0)^T$, $\mathbf{z} = (0,0,1)^T$ and $s_3$ is a constant in the $h\to 0$ limit. Hence, it suffices to study the reduced LdG tensor
 as given below:
\begin{gather}
\mathbf{Q} = \begin{bmatrix}
s_1 & s_2 \\
s_2& -s_1\\
\end{bmatrix},
\end{gather}
Here, we use similar ideas of dimension reduction to study confined BCN systems on a square domain $\Omega = [0,\lambda]^2$, where $\lambda$ is the square edge length. We describe the GS and ES-induced ordering by two reduced LdG tensors: $\Q_g$ and $\Q_c$ as described below:
\begin{gather}
\mathbf{Q}_g = \begin{bmatrix}
q_1 & q_2 \\
q_2& -q_1\\
\end{bmatrix},
\mathbf{Q}_c = \begin{bmatrix}
p_1 & p_2 \\
p_2& -p_1\\
\end{bmatrix},
\e = (e_1,e_2)^T,\nonumber
\end{gather} where $q_1, q_2,p_1,p_2$ only depend on $x$ and $y$, and $\e$ is a constant vector. Recall that in \eqref{eq:ng_2D} and \eqref{eq:nc_2D}, we have
$\Q_g = \sqrt{2} S_g \left(\n_g \otimes \n_g - \frac{\mathbf{I}_2}{2} \right)$ and $\n_g = (\cos \theta, \sin \theta)$  for some director angle $\theta$ in the plane, and analogously $\Q_c = \sqrt{2} S_c \left(\n_c \otimes \n_c - \frac{\mathbf{I}_2}{2} \right)$ and $\n_c = (\cos \phi, \sin \phi)$  for some director angle $\phi$ in the plane. In other words, 
\begin{gather}
q_1 = S_g \cos 2\theta/\sqrt{2},\ q_2 = S_g\sin 2 \theta/\sqrt{2},\nonumber\\
p_1 = S_c \cos 2\phi/\sqrt{2},\ p_2 = S_c\sin 2 \phi/\sqrt{2}. \label{eq:SgScthetaphi}
\end{gather}

Using the re-scaling, $\tilde{x} = \frac{x}{\lambda}$ and the assumption $K_g = K_c$, the non-dimensionalised free energy can be written as (refer to \eqref{eq:F})
\begin{align}
&\tilde{F}(\Q_g,\Q_c) = \int_{\tilde{\Omega}} 2(|\tilde{\nabla} q_1 |^2 + |\tilde{\nabla} q_2 |^2 + |\tilde{\nabla} p_1 |^2 + |\tilde{\nabla} p_2 |^2)\nonumber\\
   & + 2\tilde{\lambda}^2 \left(A(q_1^2+q_2^2) + C(q_1^2+q_2^2)^2\right.\nonumber\\
   &\left.-\sqrt{2}E(q_1e_1^2 + 2q_2 e_1e_2 - q_1e_2^2)\right.\nonumber\\
   &\left.+ M(p_1^2+p_2^2) + N(p_1^2+p_2^2)^2-2D(p_1q_1+p_2q_2)\right.\nonumber\\
   &\left.-\sqrt{2}PE(p_1e_1^2 + 2p_2 e_1e_2 - p_1e_2^2)\right)d\tilde{x},\label{eq:2D_energy}
\end{align}
 where $\tilde{\Omega} = [0,1]^2$ and $\tilde{\lambda}^2 = C_g\lambda^2/(2K_g)$.
The critical points of \eqref{eq:2D_energy} are solutions of the corresponding Euler-Lagrange equations:
\begin{equation}
\begin{aligned}
\tilde{\Delta}q_1 &= \tilde{\lambda}^2 \left(q_1\left(2C(q_1^2+q_2^2)+A\right)-Dp_1-\frac{\sqrt{2}}{2}E(e_1^2-e_2^2)\right),\\
\tilde{\Delta}q_2 &= \tilde{\lambda}^2 \left(q_2\left(2C(q_1^2+q_2^2)+A\right)-Dp_2-\sqrt{2}Ee_1e_2\right),\\
\tilde{\Delta}p_1 &= \tilde{\lambda}^2 \left(p_1\left(2N(p_1^2+p_2^2)+M\right)-Dq_1-\frac{\sqrt{2}}{2}PE(e_1^2-e_2^2)\right),\\
\tilde{\Delta}p_2 &= \tilde{\lambda}^2 \left(p_2\left(2N(p_1^2+p_2^2)+M\right)-Dq_2-\sqrt{2}PEe_1e_2\right),\label{eq:EL}
\end{aligned}
\end{equation}
with Dirichlet boundary conditions
\begin{align}
&q_1 = S_g^*/\sqrt{2}\ on\ y=0, 1,\ q_1 = -S_g^*/\sqrt{2}\  on\ x=0, 1\label{eq:bc1}\\
&q_2 = p_1=p_2=0\ on\ x=0, 1,\ and\ y=0, 1.\label{eq:bc2}
\end{align}
Here $S_g^*$ is the positive solution of \eqref{eq:Sg_Sc1}-\eqref{eq:Sg_Sc2} with $B = 0$, minimising bulk energy density $f_b$ in \eqref{eq:fb}. We impose tangential boundary conditions on $\Q_g$, i.e., $\n_g = (\pm 1, 0)$ on $y=0,1$ and $\n_g = (0, \pm 1)$ on $x=0,1$. 
These translate to conflicting boundary conditions for $q_1$ on the horizontal and vertical edges. To deal with this, there must be an interior nodal line with $q_1=0$, with either $S_g=0$ or with  $\n_g = (1/\sqrt{2},\pm 1/\sqrt{2})$ (captured by $\theta = \pm\frac{\pi}{4}$).
For $\Q_c$, we impose $S_c = 0$ on the square edges as in \cite{madhusudana2017two,patranabish2019one,mondal2025}, to exclude cybotactic clusters on the square edges. 

\begin{figure*}
    \centering
    \includegraphics[width=1.0\textwidth]{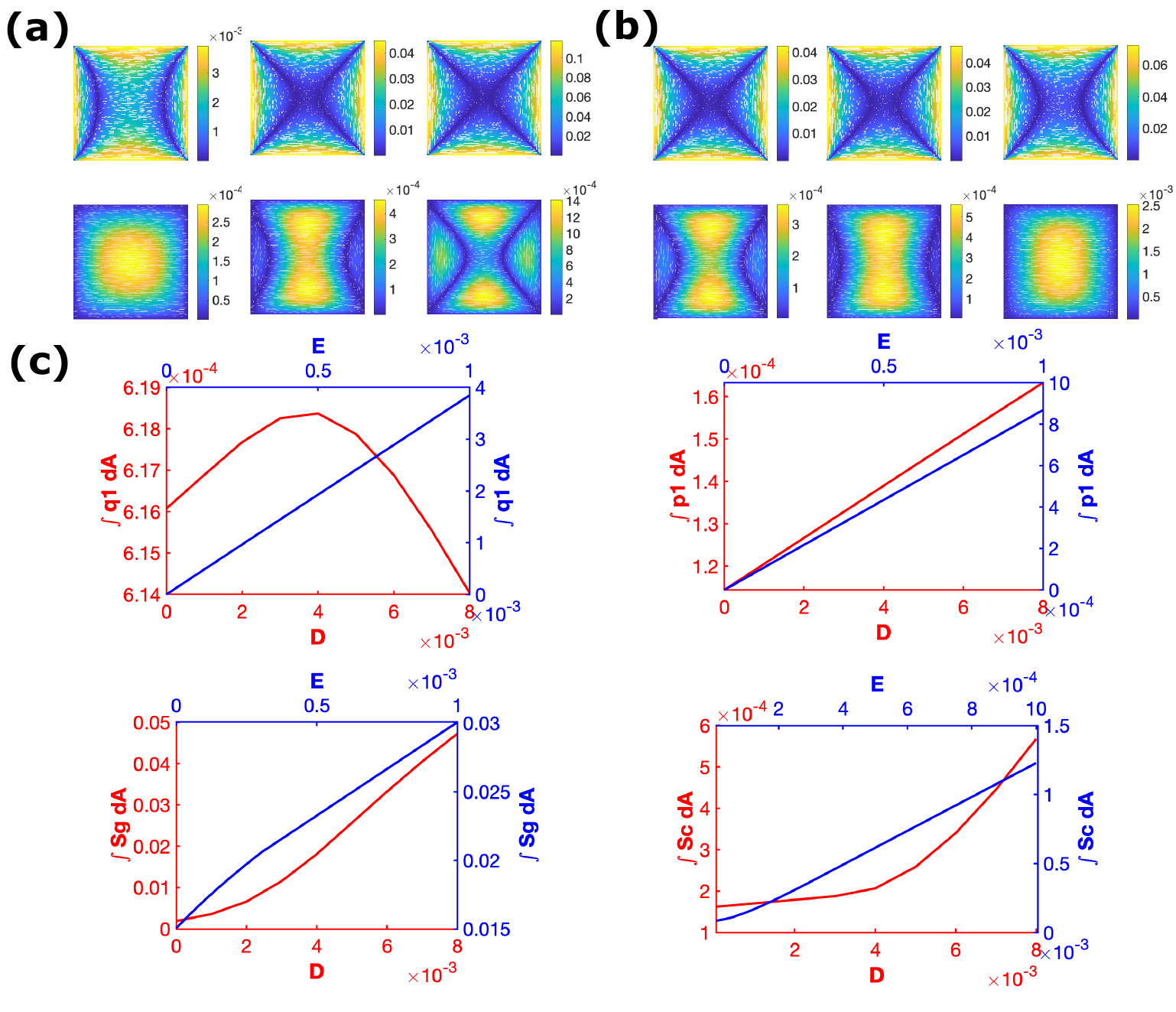}\\
    \caption{Solution pairs $(\Q_g,\Q_c)$ of \eqref{eq:EL} with $\tilde{\lambda}^2 =200$, $A=0.04$ and $\e = (1,0)$. In panel (a), we maintain $E = 1.6\times 10^{-4}$, while varying $D$ with values $0$, $4\times 10^{-3}$, and $8\times 10^{-3}$ from left to right. In panel (b), we maintain $D = 4\times 10^{-3}$, while varying $E$ with values $10^{-4}$, $2\times 10^{-4}$, and $1\times 10^{-3}$ from left to right. $\Q_g$ and $\Q_c$ are shown in the top and the bottom row, respectively.
    The colour bar labels the order parameter $S_g=\sqrt{2(q_1^2 + q_2^2)}$, $S_c=\sqrt{2(p_1^2 + p_2^2)}$. The white lines label the director of $\n_i$, $i = g,c$ in \eqref{eq:ng_2D} and \eqref{eq:nc_2D}. This plotting method also applies to Fig. \ref{fg:large_DE}-Fig. \ref{fg:D}. (c) The plots of symmetry measurements $\int q_1 dA$ and $\int p_1 dA$, the average orientational order $\int S_g dA$ and $\int S_c dA$ vs $D$ (red) or $E$ (blue). 
    }
  \label{fg:small_DE}
\end{figure*}

\begin{figure*}
    \centering
    \includegraphics[width=0.5\textwidth]{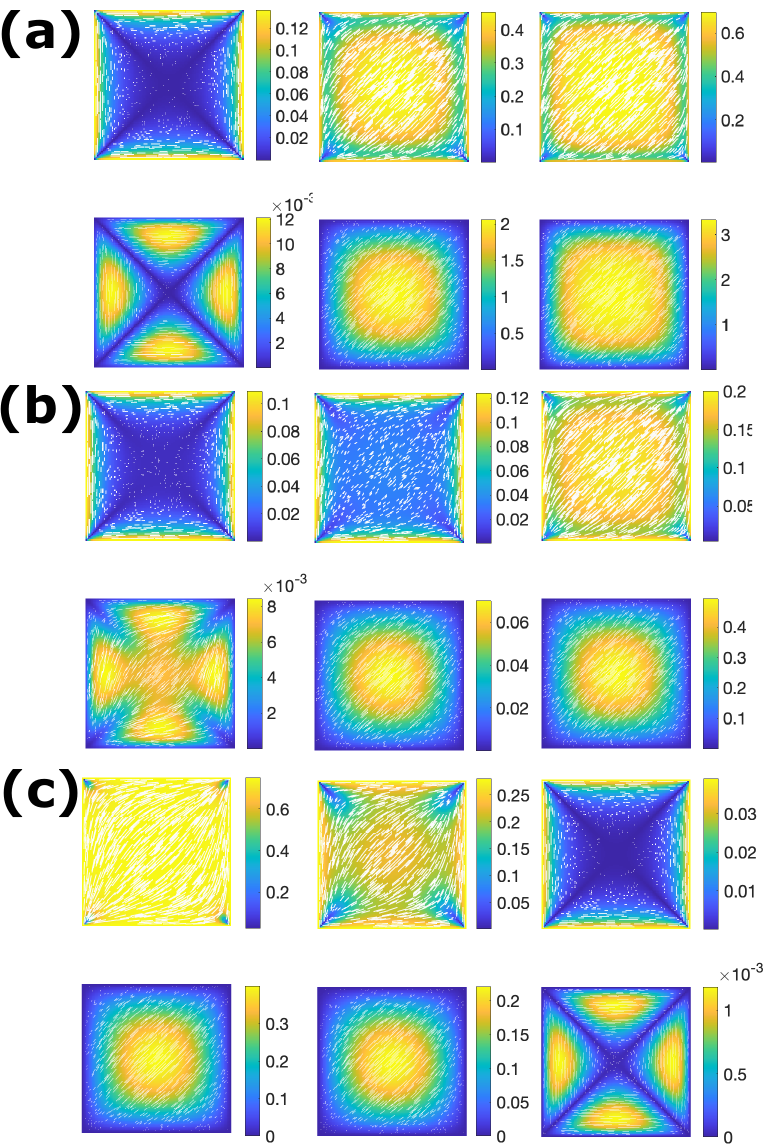}\\
    \caption{Solution pairs $(\Q_g,\Q_c)$ of \eqref{eq:EL} with $\tilde{\lambda}^2 =2500$, $A=0.04$ and $\e = (1/\sqrt{2},1/\sqrt{2})$. In panel (a), we maintain $E = 0$, while varying $D$ with values $10^{-2}$, $5\times 10^{-2}$, and $10^{-1}$ from left to right. In panel (b), we maintain $D = 8\times 10^{-3}$, while varying $E$ with values $10^{-4}$, $10^{-3}$, and $10^{-2}$ from left to right.  In panel (c), we maintain $D = 8\times 10^{-3}$, $E = 0$, while varying $A$ with values $-0.5$, $-0.04$, and $0.4$ from left to right.
    }
  \label{fg:large_DE}
\end{figure*}

In Figs. \ref{fg:small_DE} and \ref{fg:large_DE}, we focus on the high temperature case, $A = 0.04$, and study the effects of coupling parameter $D$ and external field $E$ on the energy minimisers in \eqref{eq:2D_energy}. 

For the small domain case in Figure~\ref{fg:small_DE}, we recover the WORS (Well Order Reconstruction Solution) \cite{kralj2014order} where $S_g=0$ on the two square diagonals, and the BD (Bent Director/Boundary Distortion) solution \cite{wang2019order} where $S_g=0$ on two curves localised near parallel square edges. We find that the WORS is stabilised by increasing $D$ for positive $A$. As $D$ increases in Fig. \ref{fg:small_DE}(a), the average orientational order $\int S_g dA$ and $\int S_c dA$ increases (Fig. \ref{fg:small_DE}(c)), corresponding to the fact that $S_g$ and $S_c$ increase with increasing $D$ for bulk energy minimisers in Fig \ref{fg:Fig2}. In our previous work \cite{robinson2017molecular, han2021solution}, as the average orientational order increases, the nodal lines of WORS deform from the diagonal lines to shorter lines along opposite square edges (BD). However, we observe the opposite trend here. The boundary condition of $\Q_g$ depends on $S_g^*$ in \eqref{eq:bc1}, which is an increasing function of $D$, leading to stronger constraining effects which stabilise symmetric structures like WORS.
The symmetry of a profile is measured by $\int q_1 dA$ and $\int p_1 dA$ in (Fig. \ref{fg:small_DE}(c)). 
The integral $\int q_1 dA$ initially increases as $D$ increases due to enhanced ordering and then decreases, indicating the increasing symmetry of the $\Q_g$ profile. Since $\tilde{\lambda}^2$ is relatively small, $\Q_c$ is significantly affected by the isotropic boundary condition in \eqref{eq:bc2}, so that $S_c$ is around $10^{-4}$ and is much smaller than its value in the homogeneous case in Fig. \ref{fg:Fig2}. Hence, $\Q_c$ has negligible impact on $\Q_g$ and the system is effectively decoupled.  

As $E$ increases in (Fig. \ref{fg:small_DE}(b)) for a fixed $D$, 
the effects of the boundary conditions and $S_g^*$ do not change significantly from left to right. 
Hence, consistent with our previous work, the interior ordering increases, the nodal lines of $\Q_g$ become further apart and the profile is closer to the BD profile.  All the four measurements, the average orientational order $\int S_g dA$ and $\int S_c dA$ and the symmetry measurements $\int q_1 dA$ and $\int p_1 dA$, reflect the transition from a WORS to a BD-type profile with increasing $E$. 

In Figure~\ref{fg:large_DE}, we consider a large domain $\tilde{\lambda}^2 = 2500$. For large $\tilde{\lambda}$, we expect energy minimisers of \eqref{eq:2D_energy} to converge to minimisers of the bulk energy uniformly, almost everywhere away from defects. In other words, we expect $S_g$ to converge almost uniformly to $S_g^*$ in the square interior, for sufficiently large $\tilde{\lambda}$. The bulk energy minimisers for $A=0.04$ are plotted as a function of $D$ in Figure~\ref{fg:Fig2}, with $E=0$. We note that $D$ is very large in Figure~\ref{fg:large_DE} and the corresponding value of $S_g^*$ is also expected to be large for $D> 10^{-2}$. Hence, nodal lines are expensive and the energy-minimising $\Q_g$ profile mediates between the conflicting boundary conditions by means of a diagonal profile with $\n_g = \left(1/\sqrt{2}, \pm 1/\sqrt{2}\right)$ for $D=5\times 10^{-2}$ and $D=10^{-1}$. For $D=10^{-2}$, there is competition between the elastic distortion effects and the interior ordering promoted by large $\tilde{\lambda}$ and $D$. It seems that the energy-minimising $\Q_g$ profile prefers a WORS-type profile for $D=10^{-2}$ with diagonal lines of low order.  The cybotactic clusters become pronounced in the square interior with large values of $|\Q_c|^2$ since the effects of the isotropic boundary conditions are dominated by the effects of the bulk energy minimiser in \eqref{eq:fb} (with $B=0$) in the interior, for large $\tilde{\lambda}$ and large $D$. We note that one cannot achieve an ordered diagonal solution with $D=0$, $\tilde{\lambda}^2 = 2500$ and $A=0.04$, but the diagonal solution can be stabilised by large $D$ at high temperatures.

In Figure~\ref{fg:large_DE}(b), we fix $D$ and $A = 0.04$ and the direction of the external field to be in the diagonal direction, and increase the strength of the external field from left to right. The effect of increasing $E$ is analogous to the effect of increasing $D$ for fixed $A=0.04$; $S_g^*$ is an increasing function of $E$ for fixed $D$ and the order parameters of the energy minimising $(\Q_g, \Q_c)$-profiles converge almost uniformly to the bulk energy minimisers $(S_g^*, S_c^*)$ in \eqref{eq:fb}, for sufficiently large $\tilde{\lambda}$. Hence, we note a transition from a WORS-type profile to a diagonal $\Q_g$ solution with increasing $E$, for a fixed $D$ at $A=0.04$. Analogously, the cybotactic clustering is also enhanced in the interior with increasing $E$ or increasing $D$, for $A=0.04$, and this manifests in enhanced values of $|\Q_c|^2$ in the square interior. 
We expect the qualitative conclusions to carry over to arbitrary high temperatures captured by fixed $A>0$, while  we recover only the isotropic phase at high enough temperatures for fixed $D$ or $E$. 

In Figure~\ref{fg:large_DE}(c), we fix $D$ and $E = 0$, and increase the temperature $A$ from left to right. The effect of increasing $A$ is inverse to the effect of increasing $D$ or $E$. According to the results in \cite{madhusudana2017two}, $S_g^*$ and $S_c^*$ are decreasing functions of $A$. Hence, we note a transition from a diagonal $\Q_g$ solution to a WORS-type profile with increasing $A$. The cybotactic clustering is weakened in the interior with increasing $A$, and this manifests in reduced values of $|\Q_c|^2$ in the square interior.

\begin{figure*}
    \centering
    \includegraphics[width=0.8\textwidth]{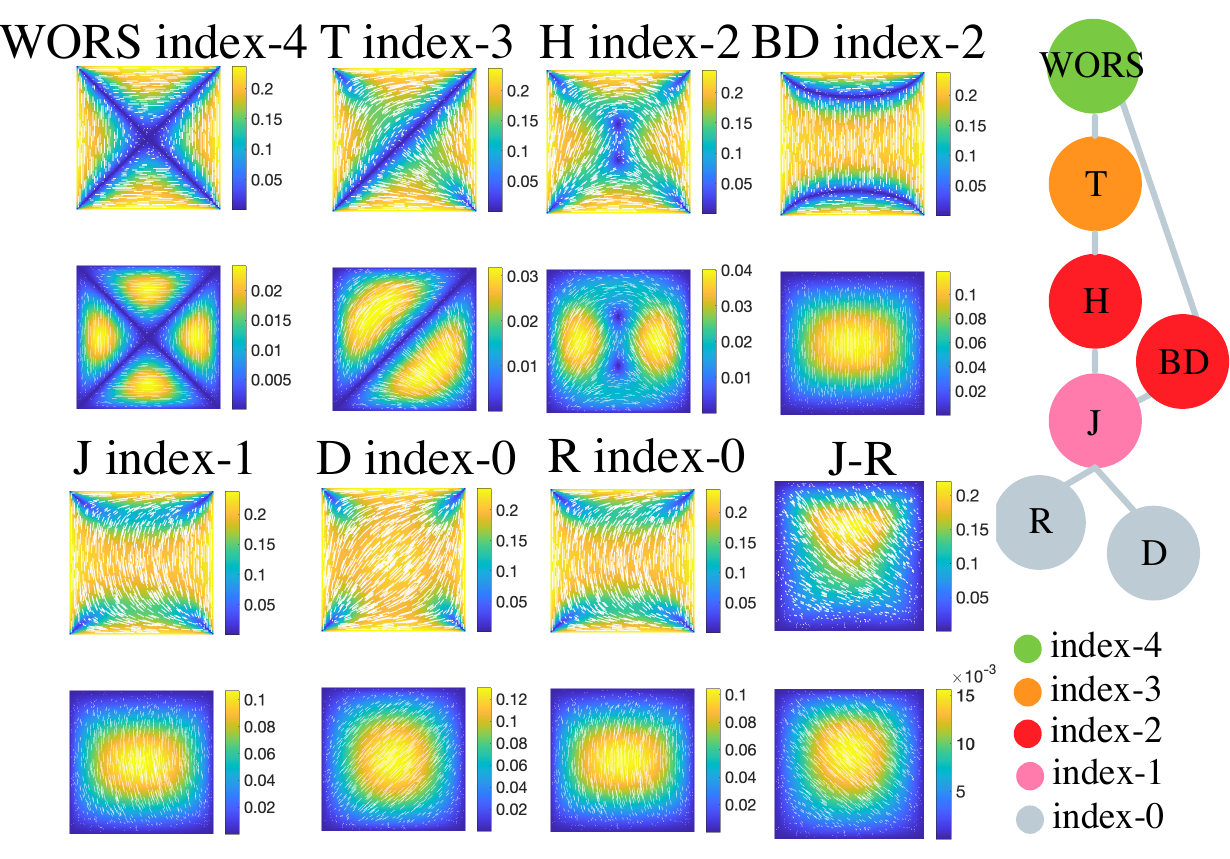}
    \caption{Left panel: solution pairs $(\Q_g,\Q_c)$ of \eqref{eq:EL}  with zero external field ($E= 0$) and $\tilde{\lambda}^2 =2500$, $A=-0.04$, $D = 4\times 10^{-3}$, $M = 9.7\times 10^{-5}$, $N = 1.78\times 10^{-3}$ and the difference between J and R solutions. Right panel: the connectivity between different solution pairs.
    }
  \label{fg:SL}
\end{figure*}
Figures~\ref{fg:small_DE} and \ref{fg:large_DE}(a-b) focus on the high-temperature regime $A=0.04$. 
In Fig. \ref{fg:SL}, we focus on computing solutions of the Euler-Lagrange equations \eqref{eq:EL} in the low temperature regime $A=-0.04$, where the global bulk energy minimiser is always the positive solution branch, $(S_g^*, S_c^*)$ in Fig~\ref{fg:Fig2}, with and without an external field. We work with a large domain $\tilde{\lambda}^2 = 2500$ and relatively weak coupling $D = 4\times 10^{-3}$ in Figure \ref{fg:SL}. 
We note that $\tilde{\lambda}^2 = 2500$ so that we expect energy minimisers to approach $(S_g^*, S_c^*)$ in the square interior (the bulk energy dominates the elastic energy for large $\tilde{\lambda}$). Further, the values $M = 9.7\times 10^{-5}$, $N = 1.78\times 10^{-3}$ are quite small and in this respect, we expect $\Q_c$ to be tailored by $\Q_g$, i.e., the coupling term promotes co-alignment of $\n_g$ and $\n_c$ and the coupling term dominates the bulk potential of $\Q_c$ (parameterised by $M$ and $N$). As such, we largely recover the $\Q_g$ profiles as reported in the reduced LdG study (with $D=0$) in  \cite{yin2020construction}. The $\n_c$ profiles follow the $\n_g$ profiles with enhanced interior cybotactic clustering tailored by $S_c^*$.
For example, we find the following solutions of \eqref{eq:EL} or critical points of \eqref{eq:2D_energy}: the T solution which is disordered along a single diagonal line; the H solution, featuring a +1/2 and a -1/2 defect; the J solution, which has lower order near the top boundary than R;  R (rotated solution) for which $\n_g$ rotates by $\pi$ radians between a pair of opposite square edges (Fig. \ref{fg:SL}(a)). We follow the terminology used in \cite{yin2020construction}.
The stability of a critical point can be measured in terms of its Morse index \cite{milnor2016morse}, i.e., the number of negative eigenvalues of the second variation of the generalised energy \eqref{eq:2D_energy} evaluated at the critical point or equivalently, the number of unstable eigendirections for a critical point, $(\Q_g, \Q_c)$ of \eqref{eq:2D_energy}. The index of these solutions is almost the same as in the reduced LdG model with $D=0$ except that the index of the WORS is $5$ with $D=0$. When $D=0$, the WORS has an unstable eigenvector corresponding to a negative eigenvalue close to zero. This negative eigenvalue changes sign when $D>0$, making the WORS index-$4$ with weak coupling. The unstable critical points with positive Morse index are not experimentally observable but they play a crucial role in transition pathways between stable critical points and the selection of the stable solution for multistable systems. For example, index-$1$ unstable critical points are often referred to as transition states and can be experimentally observed during switching processes. Moreover, unstable critical points can be stabilised by external controls.
 For example, the unstable BD configuration for $D = E = 0$ becomes stable with the constraint $q_2=p_2\equiv 0$ or under an applied electric field, as shown in Fig.~\ref{fg:small_DE}.

We also use HiOSD dynamics \cite{yin2019high} in Sec. \ref{sc:nm} to construct pathways between the distinct critical points in Figure~\ref{fg:SL} with the WORS as the parent state. If we are able to find a critical point B by perturbing the critical point A, by following the HiOSD dynamics, we say that the critical points A and B are connected.
These pathways/connections in Fig. \ref{fg:SL}(b) can provide guidance on how to effectively manipulate the defects, directors and cluster properties in confined BCNs.

\begin{figure*}
    \centering
    \includegraphics[width=0.8\textwidth]{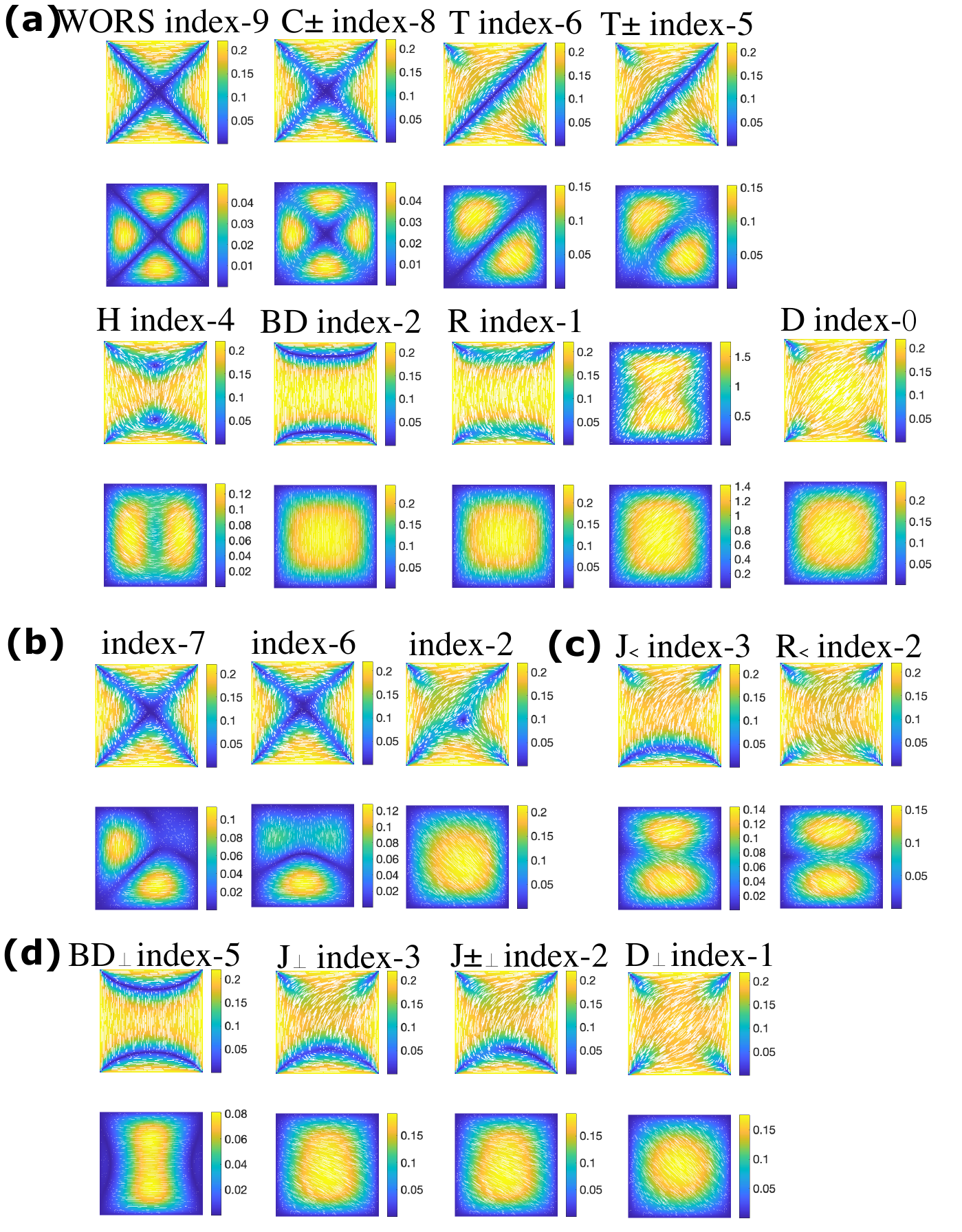}
    \caption{Solution pairs, $(\Q_g,\Q_c)$ of \eqref{eq:EL} with $E= 0$, $\tilde{\lambda}^2 =2500$, $A=-0.04$, $D = 4\times 10^{-3}$, $M = -200\times(9.7\times 10^{-5})$, $N = 200\times (1.78\times 10^{-3})$. (a) Solutions with $\Q_g$ profiles reported in \cite{yin2020construction} and $\n_c$, $\n_g$ co-aligned. The unstable mode associated with the R state is shown beside it. (b) New solutions with novel $\Q_g$ and $\Q_c$ profiles. (c) Solutions J$_<$ and R$_<$ for which $\n_c$ tilts upwards to the right in the upper part and tilts downwards to the right in the lower part. (d) Solutions BD$_\perp$, J$_\perp$, J$_{\pm\perp}$, D$_\perp$ with orthogonal $\n_c$ and $\n_g$. 
    }
 \label{fg:SL_nM}
\end{figure*}
In Fig. \ref{fg:SL}, the values of $M$ and $N$ are small so that the $\Q_g$ profiles tailor the $\Q_c$ profiles, i.e., $\n_c$ co-aligns with $\n_g$ in the solutions plotted in Figure~\ref{fg:SL}. In Fig. \ref{fg:SL_nM}, we select a negative value of $M$ (this further enhances the value of $S_c^*$) and take $M$ and $N$ to be $200$ times larger in magnitude than the values in Figure~\ref{fg:SL}. We have no physical motivation for these choices but simply want to investigate the effects of $M$ and $N$ on the solution landscapes and how this can be used to yield greater autonomy to the $\Q_c$ profiles. This can necessarily deepen our understanding of the generalised model and its implications for BCN systems.

We have three notable observations. 
(i) As shown in Fig. \ref{fg:SL_nM}(a), there are some $\Q_g$-solutions similar to those reported in \cite{yin2020construction}: index-$9$ WORS, index-$8$ C$\pm$, index-$6$ T, index-$5$ T$_{\pm}$, index-$4$ H, index-$2$ BD, index-$1$ R, index-$0$ D. We note that the increased values of $M$ and $N$ seem to increase the Morse index of the $\Q_g$ solutions compared to Figure~\ref{fg:SL}. Interestingly, the rotated (R) solution—usually stable—is now destabilized in this regime. The $\Q_c$ profiles are more autonomous compared to Figure~\ref{fg:SL}.
Since the boundary condition for $\mathbf{Q}_c$ is set to zero, it is not constrained by the tangential anchoring condition.  
Therefore, compared with Fig.~\ref{fg:SL}, the $\mathbf{Q}_c$ profile of the R state in Fig.~\ref{fg:SL_nM} has less bending and closely resembles the $\Q_c$ profile of a BD solution, which is unstable in Figs. \ref{fg:SL} and \ref{fg:SL_nM}. 
We speculate this generates more unstable directions for the solution pair, $(\Q_g, \Q_c)$, and consequently, increases the corresponding Morse index. We also note that $\n_c$ and $\n_g$ are largely co-aligned for these examples. 

(ii) There are at least three novel solution pairs in Fig. \ref{fg:SL_nM}(b): the corresponding  $\Q_g$ profiles are not attainable with $D=0$ or are not reported in the existing literature. Two novel solutions are connected to the WORS; the cross-shaped defect still exists but the centre is shifted along the diagonal or $x=0.5$. The third novel solution is connected to the T solution with an interior $-1/2$ defect. The solutions in Fig. \ref{fg:SL_nM}(c) have familiar $\Q_g$ profiles but relatively novel $\Q_c$ profiles: the $\n_c$ profile tilts upwards to the right in the upper part, tilts downwards to the right in the lower part  and is roughly coaligned with $\n_g$.

(iii) We surprisingly find a set of solutions labelled as index-$5$ BD$_{\perp}$, index-$3$ J$_{\perp}$,  index-$5$ J$_{\pm\perp}$ and index-$1$ D$_{\perp}$, for which the corresponding $\n_g$ and $\n_c$ are almost perpendicular to each other (Fig. \ref{fg:SL_nM}(d)). 

\begin{figure*}
    \centering
    \includegraphics[width=0.7\textwidth]{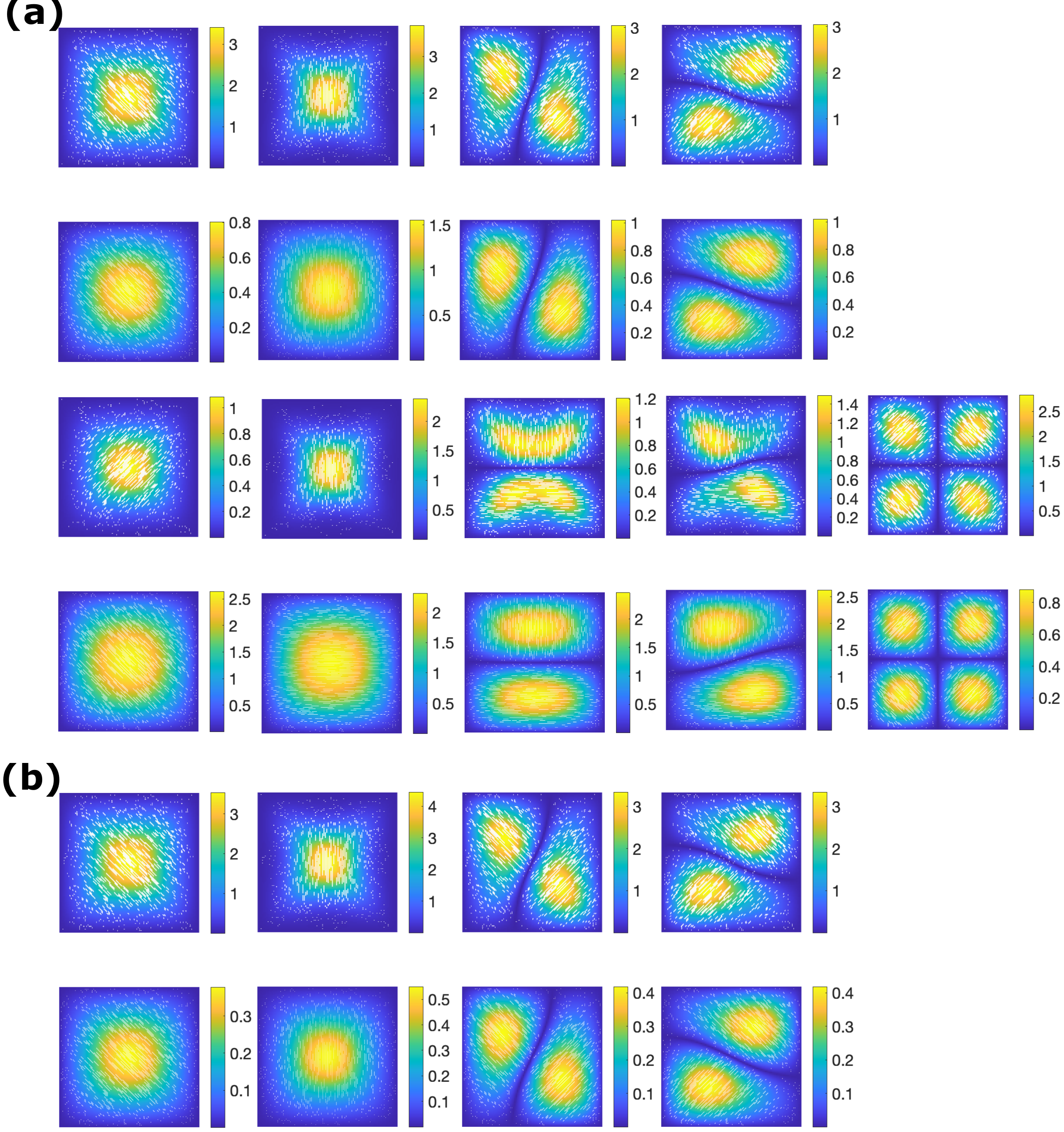}
    \caption{The unstable directions of (a) the index‑9 WORS with \(M = -200\times(9.7\times 10^{-5})\)
and \(N = 200\times (1.78\times 10^{-3})\) in Fig.~\ref{fg:SL_nM} and (b) the index‑4 WORS with \(M = 9.7\times 10^{-5}\)
and \(N = 1.78\times 10^{-3}\) in Fig.~\ref{fg:SL}.
The unstable directions correspond, from left to right, top to bottom, to the
negative eigenvalues \(\lambda_k\) introduced in Sec.~\ref{sc:HiOSD},
arranged in ascending order. The last plot in the sequence is therefore associated
with the largest (least negative) eigenvalue, i.e., the one with the smallest magnitude.}
  \label{fg:np}
\end{figure*}

We take the WORS solution---the configuration with the highest Morse index in our manuscript---as an example to illustrate how $\mathbf{Q}_c$ generates more unstable directions. 
In Fig.~\ref{fg:np}(a), the first eigenvector drives the WORS solution towards the D solution, 
as the interior order is enhanced at the center with orientations along a diagonal. 
The second eigenvector drives WORS towards the R, BD, and H states, 
where the perturbation enhances the order at the center with directions parallel to one edge of the square domain. 
The third and fourth eigenvectors are degenerate, corresponding to the formation of the $T$ and $T_{\pm}$ states. 
Their linear combinations strengthen the order on both sides of the diagonal: 
one side is oriented nearly perpendicular to the diagonal while the other is almost parallel to the diagonal. 
The first four eigenvectors of the index‑9 WORS are similar to those of the index‑4 WORS shown in Fig.~\ref{fg:np}(b). Index-9 WORS has five more unstable directions than index-4 WORS.
The combination of the seventh and eighth degenerate unstable directions leads to the index-6 and index-7 critical states in Fig.~\ref{fg:SL_nM}(b). 
The ninth unstable direction corresponds to the $C_{\pm}$ states, 
where the order is enhanced midway along the lines connecting the midpoint of the square to its four vertices, 
driving the formation of the $\pm$1 defects at the center. 
For the eigenvectors discussed above, the spatial distributions of $\n_g$ and $\n_c$, $S_g$ and $S_c$ 
in the $\mathbf{Q}_g$ and $\mathbf{Q}_c$ profiles are consistent with each other. 
However, for the fifth and sixth eigenvectors, the orientations in the $\mathbf{Q}_g$ and $\mathbf{Q}_c$ profiles 
are almost perpendicular to each other, indicating the influence of $\mathbf{Q}_c$ 
on the instability of solutions. 
The fifth unstable direction corresponds to the $D_{\perp}$ states, 
while the sixth unstable direction corresponds to the $BD_{\perp}$, $J_{\perp}$, and $J_{\pm\perp}$ states in Fig.~\ref{fg:SL_nM}(d).

\begin{figure*}
    \centering
    \includegraphics[width=\textwidth]{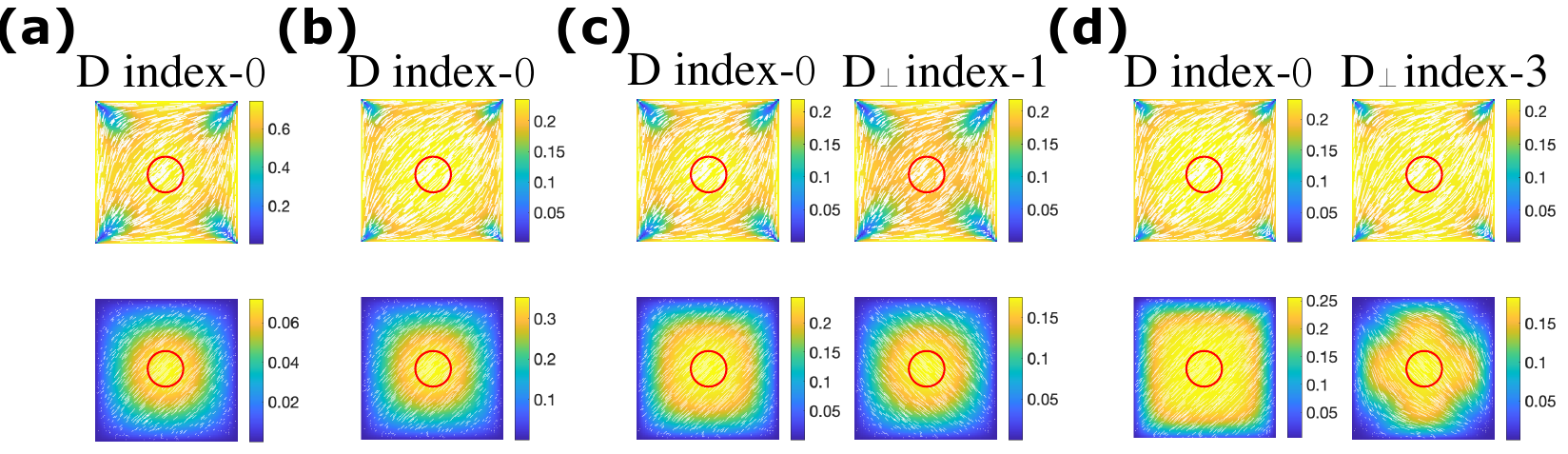}
    \caption{Solution pairs $(\Q_g,\Q_c)$ of \eqref{eq:EL} with $E= 0$, $A=-0.04$ and $D = 4\times 10^{-3}$. We set $M = 9.7\times 10^{-5}$, $N = 1.78\times 10^{-3}$ and $\tilde{\lambda}^2 =2500$ for (a), $\tilde{\lambda}^2 =5000$ for (b). We set $M = -200\times(9.7\times 10^{-5})$, $N = 200\times(1.78\times 10^{-3})$ and $\tilde{\lambda}^2 =2500$ for (c), $\tilde{\lambda}^2 =5000$ for (d). The area encircled by the red circle represents $\Omega_{\parallel\perp}$.
    }
  \label{fg:D}
\end{figure*}

Here, we provide a heuristic explanation for the emergence of critical points with orthogonal $\n_g$ and $\n_c$; see (Figs. \ref{fg:SL_nM} and \ref{fg:D}). In the absence of defects with non-zero $\Q_g$ and $\Q_c$, the variables $(q_1,q_2,p_1,p_2)$ can be written as:
\begin{gather}
q_1 = s\cos(2\theta),\ q_2 = s\sin(2\theta),\nonumber\\ p_1 = p\cos(2\phi),\ p_2 = p\sin(2\phi),\label{eq:spthetaphi}
\end{gather}
where $s = \sqrt{q_1^2+q_2^2}\geq 0$ and $p=\sqrt{p_1^2+p_2^2}\geq 0$. The relations between $S_g,S_c$ in \eqref{eq:SgScthetaphi} and $s,p$ are $s = S_g/\sqrt{2}$ and $p =S_c/\sqrt{2}$.  
Substituting the expressions in \eqref{eq:spthetaphi}
into \eqref{eq:2D_energy}, the energy can be written in terms of $(s,p,\theta,\phi)$ as given below:
\begin{equation}
\begin{aligned}
&\tilde{F}(s,p,\theta,\phi) = 2(4{s}^2|\nabla \theta|^2 + 4p^2|\nabla\phi|^2 +  |\nabla s|^2 +  |\nabla p|^2) \\
&+ 2\tilde{\lambda}^2(As^2 + Cs^4 + Mp^2 + Np^4 - 2D(sp\cos(2(\theta-\phi)))).\nonumber
\end{aligned}
\end{equation}
The corresponding Euler-Lagrange equations are:
\begin{multline}
-\Delta s + 4s|\nabla\theta|^2 + \tilde{\lambda}^2 (As + 2Cs^3\\
- D p\cos(2(\theta-\phi))) = 0,\label{eq:EL_s}
\end{multline}
\begin{multline}
-\Delta p + 4p|\nabla\phi|^2 + \tilde{\lambda}^2 (Mp + 2Np^3 \\
- D s\cos(2(\theta-\phi))) = 0,\label{eq:EL_p}
\end{multline}
\begin{align}
-2s\Delta \theta + \tilde{\lambda}^2 D s p \sin(2(\theta-\phi)) = 0,\label{eq:EL_theta}\\
-2p\Delta \phi - \tilde{\lambda}^2 D s p \sin(2(\theta-\phi)) = 0,\label{eq:EL_phi}
\end{align}

We define a defect-free region $\Omega_{\parallel\perp}$ (see Fig. \ref{fg:D}) where $\n_g$ and $\n_c$ can be either parallel or perpendicular to each other, in paired critical points, i.e.,
\begin{equation} \label{eq:thetaphizero}
\sin(2(\theta-\phi)) = 0, i.e., \phi = \theta+\frac{k\pi}{2}, k = 0,1,\cdots.
\end{equation}
In the following, we take the diagonal states D (with co-aligned $\n_g$ and $\n_c$ in $\Omega_{\parallel\perp}$) and the paired critical state D$_{\perp}$ (with $\n_g \cdot \n_c =0$ in $\Omega_{\parallel\perp}$) in Fig. \ref{fg:D}, the paired critical BD and BD$_{\perp}$ in Fig. \ref{fg:SL_nM} as examples. 

We first note that, for both types of critical points—those with parallel and those with orthogonal $\mathbf{n}_g$ and $\mathbf{n}_c$—the conditions \eqref{eq:EL_theta} and \eqref{eq:EL_phi} can be satisfied at least on $\Omega_{\parallel\perp}$.
For D and D$_{\perp}$, the associated $\Omega_{\parallel\perp}$ is located around the square centre in Figure~\ref{fg:D}.
For BD and BD$_{\perp}$, the corresponding $\Omega_{\parallel\perp}$ is also away from the square edges, the defect lines and localised near the square centre.
Substituting $\sin(2(\theta-\phi)) = 0$ into the EL equations above, the equations \eqref{eq:EL_theta}-\eqref{eq:EL_phi} reduce to
\begin{gather} \label{eq:deltazero}
\Delta\theta= \Delta\phi  = 0,
\end{gather}
when $p>0$ and $s>0$.
Due to the symmetry of D and D$_{\perp}$, 
\begin{align*}
\theta(1/2+y,1/2+x) = \pm\pi/2 - \theta(1/2+x,1/2+y),\\
\phi(1/2+y,1/2+x) = \pm(\mp)\pi/2 - \phi(1/2+x,1/2+y),
\end{align*}
we have $\Delta\theta\vert_{(0.5,0.5)}= \Delta\phi\vert_{(0.5,0.5)}= 0$ and we deduce that $\Delta\theta\vert_{\Omega_{\parallel\perp}}\approx 0$ and $\Delta\phi\vert_{\Omega_{\parallel\perp}}\approx 0$. Similarly, $\theta$ and $\phi$ are almost constant around the square centre for the paired critical points: BD and BD$_\perp$ and hence, we deduce that $\Delta\theta\vert_{\Omega_{\parallel\perp}}\approx 0$ and $\Delta\phi\vert_{\Omega_{\parallel\perp}}\approx 0$. 

As the re-scaled domain size $\tilde{\lambda}^2\to\infty$, the polynomial terms dominate in \eqref{eq:EL_s}-\eqref{eq:EL_p} and we require (to leading order) that
\begin{align*}
As + 2Cs^3 - Dp\cos(2(\theta-\phi))= 0,\\
Mp + 2Np^3 - Ds\cos(2(\theta-\phi))= 0.
\end{align*}
Substituting $\phi = \theta + \frac{k\pi}{2}$ into the above equations,
we have
\begin{align}
As + 2Cs^3 - Dp &= 0,\ for\ even\ k,\label{eq:s1}\\
Mp + 2Np^3 - Ds &= 0,\ for\ even\ k,\label{eq:p1}\\
As + 2Cs^3 + Dp &= 0,\ for\ odd\ k,\label{eq:s2}\\
Mp + 2Np^3 + Ds &= 0,\ for\ odd\ k.\label{eq:p2}
\end{align}
If \eqref{eq:s1} and \eqref{eq:p1} admit at least one positive solution pair with $s, p>0$, then we may have a critical point $(\Q_g, \Q_c)$ with approximately co-aligned $\n_g$ and $\n_c$. If the algebraic equations \eqref{eq:s2} and \eqref{eq:p2} admit a solution with positive $s$ and $p$, then we may have a critical point with almost orthogonal GS and ES directors: $\n_g$ and $\n_c$. 
 Note that \eqref{eq:s1} and \eqref{eq:p1} are the same as \eqref{eq:Sg_Sc1} and \eqref{eq:Sg_Sc2} with $B=E=0$ and $S_g=\sqrt{2}s$, $S_c=\sqrt{2}p$.
As in the previous discussion, we can derive a ninth-degree polynomial for $s$ or $p$ from \eqref{eq:s1}-\eqref{eq:p1} or \eqref{eq:s2}-\eqref{eq:p2}. Since one of the solutions is $s = p = 0$, the degree can be reduced to eight. Although, one can numerically solve the eighth-degree polynomial with specific parameter values and check for the existence of a positive solution with $s>0$ and $p>0$, it is difficult to analyse roots of an eighth-degree polynomial systematically. 

Therefore, we assume that equations \eqref{eq:s1} and \eqref{eq:s2} (\eqref{eq:p1} and \eqref{eq:p2}) are cubic polynomials in terms of $s$ ($p$).
Then the question can be partially answered by studying the number of positive roots of the following cubic equation: 
\begin{equation}
ax^3 + cx + d = 0,
\end{equation}
and how the number and sign of the roots depend on $a$, $c$ and $d$.
\begin{itemize}
\item Case 1: $a>0$, $c>0$\\
By using Cardano's formula \cite{kurosh1972higher}, since the discriminant of the cubic is given by
\begin{equation}
\Delta = (\frac{d}{2a})^2 + (\frac{c}{3a})^3>0,
\end{equation}
we have one real solution and two complex solutions.
The real solution is given by
\begin{equation}
x = \sqrt[3]{-\frac{d}{2a} + \sqrt{(\frac{d}{2a})^2 + (\frac{c}{3a})^3}} + \sqrt[3]{-\frac{d}{2a} - \sqrt{(\frac{d}{2a})^2 + (\frac{c}{3a})^3}}.\nonumber
\end{equation}
and
\begin{gather}
x<0,\ \text{if}\ d>0,\\
x=0,\ \text{if}\ d=0,\\
x>0,\ \text{if}\ d<0.
\end{gather}

\item Case 2: $a>0$, $c<0$\\
If we choose $a>0$, $c<0$ and $d$ such that 
\begin{equation}
\Delta = \left(\frac{d}{2a} \right)^2 + \left(\frac{c}{3a} \right)^3\leq 0,
\end{equation}
then we have three degenerate or non-degenerate real roots given by \cite{selby1975crc}
\begin{gather}
x_k = 2\sqrt{-\frac{c}{3a}}\cos\left(\frac{\theta}{3} + \frac{2k\pi}{3}\right),\ k = 0,1,2,\nonumber
\end{gather}
where $\theta = arccos\left(\frac{3d}{2c}\sqrt{-\frac{c}{3a}}\right)$ and
\begin{gather}
x_1<0<x_2\leq x_0,\ \text{if}\ d>0,\\
x_1<0=x_2< x_0,\ \text{if}\ d=0,\\
x_1\leq x_2<0< x_0,\ \text{if}\ d<0.
\end{gather}
\end{itemize}

These arguments can be mapped to the parameter values in Figures~\ref{fg:SL} and \ref{fg:SL_nM}.
The choice of $A = -0.04$, $C = 0.9$, $D = 4\times 10^{-3}$ in \eqref{eq:s1} and \eqref{eq:s2} corresponds to Case 2 above.
When $M = 9.7\times 10^{-5}$, $N = 1.78\times 10^{-3}$, $D = 4\times 10^{-3}$, then the polynomials in \eqref{eq:p1} and \eqref{eq:p2} correspond to Case 1. Hence, we have a positive solution pair, $(s,p)$, only if $k$ (in $\phi = \theta + \frac{k\pi}{2}$) is even in \eqref{eq:s1}-\eqref{eq:p1}, i.e., the directors $\n_g$ and $\n_c$ are almost co-aligned on $\Omega_{\parallel\perp}$.
When $M = -200\times (9.7\times 10^{-5})$, $N = 200\times(1.78\times 10^{-3})$, $D = 4\times 10^{-3}$ as in Figure~\ref{fg:SL_nM}, the polynomials in \eqref{eq:p1} and \eqref{eq:p2} correspond to Case 2 and we always have at least one positive solution $s,p>0$, for both \eqref{eq:s1}-\eqref{eq:p1} and \eqref{eq:s2}-\eqref{eq:p2}. Hence, the directors $\n_g$ and $\n_c$ can be either co-aligned or perpendicular on $\Omega_{\parallel\perp}$, yielding paired critical points $(\Q_g, \Q_c)$. 

To summarise, we have investigated solution landscapes of prototype BCN systems within the generalised model in \eqref{eq:F}, i.e., we have studied critical points $(\Q_g, \Q_c)$ of the free energy in \eqref{eq:F}. We have focused on a two-dimensional example of BCNs on square domains, subject to experimentally relevant tangent boundary conditions for $\Q_g$. We impose isotropic boundary conditions on $\Q_c$ but other choices are possible and it is reasonable to assume that cybotactic clusters are confined to the domain interior. The essential novelty is introduced by the coupling between $\Q_g$ and $\Q_c$. This coupling promotes interior ordering and stabilises ordered GS profiles even for high temperatures, when we only expect predominantly isotropic solutions at $D=0$ for high temperatures.  In fact, for $A=0.04$, the GS-ES coupling and/or external fields can stabilise the diagonal solution, which is typically only found for low temperatures. The choice of boundary conditions can also stabilise symmetric WORS-type solutions, i.e., the coupling parameter $D$ can compete with the disordering effects of high temperature. For relatively small values of $M$ and $N$, the $\Q_c$ profiles are tailored by the $\Q_g$ profiles and there are few surprises for low temperatures. For larger values of $M$ and $N$, the $\Q_c$ profiles can be more independent of the $\Q_g$ profiles leading to non-trivial and novel effects. For example, we find novel solution pairs or novel critical points in Figure~\ref{fg:SL_nM} which are not found with $D=0$. The Morse index of critical points of \eqref{eq:F} typically increases with increasing values of $M$ and $N$, since there are new unstable directions determined by the relatively autonomous $\Q_c$ profiles. For example, the rotated $R$ critical point has index-$0$ in Figure~\ref{fg:SL} and has index-$1$ in Figure~\ref{fg:SL_nM}. Importantly, we recover critical points $(\Q_g, \Q_c)$ with almost orthogonal $\n_g$ and $\n_c$ in Figure~\ref{fg:SL_nM}. Whilst we cannot make things precise, it is possible that the optical measurements are determined by an appropriate combination of $\Q_g$ and $\Q_c$ within the generalised model in \eqref{eq:F}. If $\n_g$ and $\n_c$ are orthogonal to each other in experimentally attainable BCN profiles, then this could lead to experimentally observable macroscopic biaxiality. Of course, such arguments need much more investigation than is possible in this work.

The domains with $\tilde{\lambda}^2 = 2500$ and $5000$ correspond to experimentally relevant micron-scale areas of $(0.26 \mu m)^2$ and $(0.52 \mu m)^2$, respectively. As the domain size $\lambda^2$ increases, the defects become smaller, requiring a finer computational grid. 
This, in turn, leads to a higher computational cost, lower numerical accuracy, and greater difficulty in achieving convergence. 
As discussed in the context of the LdG theory for nematics in \cite{yin2020construction} and \cite{shi2022nematic}, computations on small domains remain relevant for large domains. The solutions of the Euler-Lagrange equations on small domains serve as excellent initial conditions for numerical continuation schemes, for large values of $\lambda^2$. Most of the solution branches, for small domains, continue to exist on larger domains although the stability properties might change, i.e., some solution branches lose stability as domain size increases. However, some solution branches retain stability or gain stability, as domain size increases or $\lambda^2$ increases. The unstable solutions also play a crucial role in the system dynamics and the selection of stable solutions for multistable systems, and hence, our results for $\lambda^2 = 2500$ (although not large enough for experiments on the several-hundred-micrometer scale), can still provide direct/indirect guidance for future experimental studies.
\section{Numerical methods}\label{sc:nm}
The numerical results of Section \ref{sc:square}, including the minimisers and critical points of the non-dimensionalised free energy in 2D setting $\tilde{F}$ in \eqref{eq:2D_energy}, are calculated by the numerical methods as follows. 
The study is implemented in Matlab R2021b .
\subsection{Spatial discrete method}
We use the ﬁnite difference method to estimate the spatial derivative on a 2D square $\tilde{\Omega} = [0,1]^2$ taking the nodes $(x_i,y_j)$, $i,j=0,1,\cdots,N_0$ with step length $h=\frac{1}{N_0}$, where
\begin{equation}
    \begin{aligned}
        0=x_0 \leqslant x_1\leqslant \cdots\leqslant x_{N-1}\leqslant x_{N}=1,x_i=ih,\\
        0=y_0\leqslant y_1\leqslant \cdots\leqslant y_{N-1}\leqslant y_{N}=1,y_j=jh.\nonumber
    \end{aligned}
\end{equation}
The five-point stencil method is used to approximate the Laplacian \(\nabla^2 \hat{\x}\) of a function \( \hat{\x}(x,y) \) at a point \((x_i,y_j)\) on the discrete grid with second-order accuracy:
\begin{align*}
\nabla^2 \hat{\x}(x_i,y_j) \approx &\left(\hat{\x}_{x_{i-1},y_j} + \hat{\x}_{x_{i+1},y_j} \right.\\
&\left.+ \hat{\x}_{x_i,y_{j-1}} + \hat{\x}_{x_i,y_{j+1}}- 4\hat{\x}_{x_i,y_j}\right)/h^2.
\end{align*}
\subsection{HiOSD method}\label{sc:HiOSD}
We use the HiOSD method \cite{yin2019high}, which is a generalization of the optimization-based shrinking dimer method \cite{zhang2016optimization} to find stable and unstable solutions of the Euler-Lagrange equations in \eqref{eq:EL} on the energy landscape of $\tilde{F}$ in \eqref{eq:2D_energy}. 

The stability of a critical point of the free energy can be measured by its Morse index \cite{milnor2016morse}.
For a non-degenerate index-$k$ saddle point $\hat{\x} = (\Q_g,\Q_c)$, the Hessian $\mathbb{H}(\hat{\x})$ has exactly $k$ negative eigenvalues $\hat{\lambda}_1\leqslant \cdots \leqslant\hat{\lambda}_k$ with corresponding unit eigenvectors $\hat{\mathbf{v}}_1,\cdots , \hat{\mathbf{v}}_k$ satisfying $\big\langle\hat{\mathbf{v}}_j, \hat{\mathbf{v}}_i\big\rangle = \delta_{ij}$, $1\leqslant i, j \leqslant k$.
Define a $k$-dimensional subspace $\hat{\mathcal{V}}=\mathrm{span}\big\{\hat{\mathbf{v}}_1,\cdots, \hat{\mathbf{v}}_k\big\}$, then $\hat {\x}$ is a local maximum on a $k$-dimensional linear manifold $\hat{\x}+\hat{\mathcal{V}}$ and a local minimum on $\hat{\x}+\hat{\mathcal{V}}^\perp$, where $\hat{\mathcal{V}}^\perp$ is the orthogonal complement space of $\hat{\mathcal{V}}$. In particular, a stable state is an index-$0$ solution. 

An index-$k$ critical point can be found by the $k$-HiOSD dynamics with a certain initial condition.
A $k$-HiOSD dynamics is a transformed gradient flow of $\x$ coupling with the search for an orthonormal basis $\mathcal{V}$ which minimises $k$ Rayleigh quotients simultaneously. 
\begin{equation}
    \left\{
    \begin{aligned}
		\dot{\x}&=- (\mathbf{I}-2\sum_{i=1}^k \mathbf{v}_i\mathbf{v}_i^T)\nabla \tilde{F}(\x), \\
		  \dot{\mathbf{v}}_i&=-   (\mathbf{I}-\mathbf{v}_i\mathbf{v}_i^T-\sum_{i=1}^{i-1}2\mathbf{v}_j\mathbf{v}_j^T)\nabla^2 \tilde{F}(\x) \mathbf{v}_i,i=1,2,\cdots,k .\\
    \end{aligned}
    \right.\nonumber
\end{equation}
To avoid evaluating the Hessian of $\tilde{F}(\x)$, we use central difference schemes for directional derivatives to approximate Hessians along $i$th dimer with length $2l$ centred at $\x$,
\begin{equation}\nonumber
\nabla\tilde{F}^2(\x)\mathbf{v}_i\approx\frac{\nabla \tilde{F}(\x+l\mathbf{v}_i)-\nabla \tilde{F}(\x-l\mathbf{v})}{2l}
\end{equation}
We use the time-discrete Euler scheme with $\Delta t = 10^{-4}$.

We obtain a critical point of the free energy if the HiOSD dynamics converges with error tolerance $|\nabla\tilde{F}(\hat{\x})|\leq 10^{-6}$.
The index of the critical point is checked by the calculation of the smallest $k+1$ eigenvalues of the Hessian $\nabla^2 \tilde{F}(\hat{\x})$ by using the simultaneous Rayleigh-quotient minimisation method \cite{longsine1980simultaneous}. 

\subsection{Algorithm for building the connectivity of critical solutions}
Following the HiOSD dynamics, we build the connectivity of critical points using two algorithms: a downward search that enables us to search for all connected lower-index saddles from an index-$k$ saddle; an upward search with a selected direction to find the higher-index saddles. The two algorithms drive the entire search to navigate up and down on the energy landscape. 
Please refer to \cite{han2021solution} for detailed procedure of downward and upward search.

\section{Conclusion} \label{sec:conclusions}

In this paper, we propose and study a generalised LdG-type model for BCN, building on the two-state model proposed by Madhusudana in \cite{madhusudana2017two}. Our model is based on two tensor order parameters: $\Q_g$ and $\Q_c$ instead of two scalar order parameters, $S_g$ and $S_c$ as in \cite{madhusudana2017two}. Our model has all the capabilities of Madhusudana's model and more --- it can account for the GS director and additional directors induced by the cybotactic clusters, spatial inhomogeneities, effects of boundary conditions, geometric frustration, defects etc. Notably, we find that the GS-ES coupling, measured in terms of $D$ in \eqref{eq:F}, can induce phase transitions at a fixed temperature. For example, we can reproduce the nematic-paranematic phase transition and the paranematic-isotropic phase transitions by reducing $D$, at a fixed high temperature. Madhusudana reports the same sequence of phase transitions with increasing temperature at fixed $D$, in 3D. We also study phase transitions in 2D although strictly 2D systems are not realistic. However, $\Q_g$ and $\Q_c$ are reduced to symmetric traceless $2 \times 2$ matrices in quasi-2D settings as in Section~\ref{sc:square} and as such, the 2D phase diagrams in Section~\ref{sec:phasetransitions} can shed light into planar ordering and disordering as a function of $D$ in such settings. The 2D phase diagrams in Section~\ref{sec:phasetransitions} certainly help us understand structural phase transitions in confined quasi-2D systems as a function of $D$. Of course, we do not have any physical insight into how to tune or manipulate $D$, except that it is an intrinsic material parameter.

Our model can be embellished to capture emergent chirality, transitions to the twist-bend and splay-bend phases and macroscopic biaxiality. This could be done by introducing elastic anisotropy into the generalised free energy, i.e., more general and higher-order terms (cubic terms such as $\Q_g \nabla \Q_g \nabla \Q_g$) in the elastic energy in \eqref{eq:F}. Inspiration can be sought from existing work on the critical role of elastic constants in the emergence of the twist-bend and splay-bend phases in BCN; see \cite{jaklirevmodernphysics2018}. However, we have limited insight into typical values for $K_c$ --- the elastic constant associated with $\Q_c$. We have employed a one-constant approximation in this paper, with $K_g = K_c$, but the precise form of the elastic energy of $\Q_c$ is an open question. Octupolar order could be introduced by adding a higher-order tensor such as $T_{ijk}$ with an appropriate physical interpretation \cite{jaklirevmodernphysics2018} but open questions remain with regards to the associated energetic contributions.


It is commendable that a simple two order parameter model in \cite{madhusudana2017two}, further generalised in our paper, is capable of capturing some complex experimental trends. This is, in itself, a strong reason for studying such models since they offer tractable approaches to complex multi-physics problem. However, they might miss subtle underpinning physics about the cybotactic clusters and can only offer a coarse-grained homogenised perspective. 
For example, our model does not have additional positional order parameters to account for density modulations within cybotactic clusters. This model does not account for the spacing between the cybotactic clusters or the Smectic-type ordering within the cybotactic clusters.
To describe these additional features, one may need to adopt a model for SmC‑type ordering, such as \cite{xia2024simple}, 
in which a density order parameter is introduced in addition to $\mathbf{Q}_c$, 
and a coupling term between the layer normal (which can be represented through the first‑ or second‑order derivatives of the density field) 
and $\mathbf{Q}_c$ is included in the free energy. 
This leads to a more complex mathematical formulation, which will be explored in our future work. Further, Madhusudana's model and our generalised model can offer no insight into how the opening angle of BCN molecules affects the macroscopic BCN phases. 
We employ a generic coupling term proportional to $\mathbf{Q}_g \cdot \mathbf{Q}_c$ as an immediate generalisation of the GS-ES coupling term in \cite{madhusudana2017two}. This is a canonical coupling term that arises in the dilute limit of suspensions, e.g., when we treat the cybotactic clusters as a dilute suspension of smectic nanoparticles or nanoclusters in a nematic GS medium, consistent with the assumptions in Madhusdana's paper \cite{canevarihomogenisation}.  
Similar coupling terms are also employed in phenomenological studies of ferronematics --- dilute suspensions of magnetic nanoparticles in nematic media \cite{hanpreferronematics}. This coupling term can miss nonlinear phenomena. For example, as the global nematic order 
$\mathbf{Q}_g$ increases, $\mathbf{Q}_c$ may exhibit nonlinear behavior such as saturation characteristics. In bent‑core nematics, the molecules are intrinsically bent and typically respond differently along the bending plane and the perpendicular plane. Possible improvements include introducing higher‑order or anisotropic couplings. Further, it is often difficult to decide on the exact form of the elastic energy associated with $\mathbf{Q}_c$ or the exact origins of the Ginzburg-Landau type potential for $\mathbf{Q}_c$ --- what is the underlying physics? Finally, how do we estimate the parameters $M, N$ and $D$ in the free energy  and relate these phenomenological parameters to physical experiments on BCNs in a reliable way? Optical measurements probably only yield a mean scalar order parameter as suggested in \cite{madhusudana2017two} but one needs quantitative information about $S_c$ to infer meaningful information about $M$, $N$ and $D$. Open questions remain about the interpretability of this model but it is undeniable that this simple two order parameter model captures complex nonlinear phenomena qualitatively and certainly gives insight into cybotactic cluster formation, location and stability in prototype confined systems, making this a worthwhile and interesting study.

\begin{acknowledgements}
YH gratefully thanks Prof. Lei Zhang and Beijing International Center for Mathematical Research of Peking University for hosting her as a Visiting Scholar. PR acknowledges support from Leverhulme Research Project Grant RPG-2021-401.
AM is supported by the University of Strathclyde New Professors Fund, the Humboldt Foundation and a Leverhulme Research Project Grant RPG-2021-401.
\end{acknowledgements}

\bibliography{main}

\end{document}